\documentclass[10pt,journal,compsoc]{IEEEtran}
\newsavebox\tempbox
\newlength\templen
\usepackage{multirow}
\usepackage{array}
\usepackage{rotating}
\usepackage{xcolor,stfloats}
\usepackage{adjustbox}
\usepackage{booktabs}
\usepackage{pifont}
\usepackage{framed}
\usepackage{listings}
\usepackage{subfigure}
\usepackage{ulem}
\usepackage{amsmath}
\usepackage{stfloats}
\usepackage[ruled,vlined,linesnumbered]{algorithm2e}
\usepackage{tikz}
\newcommand*{\circled}[1]{\lower.7ex\hbox{\tikz\draw (0pt, 0pt)%
circle (.6em) node {\makebox[1em][c]{\small #1}};}}
\SetKw{Continue}{continue}
\SetKw{Break}{break}
\SetKw{Return}{return}
\newcommand{\cmark}{\ding{51}}%
\newcommand{\xmark}{\ding{55}}%

\newcommand{\ignore}[1]{}

\usepackage{tikz}
\usepackage{amsmath}
\usepackage{url}

\ifCLASSOPTIONcompsoc
  \usepackage[nocompress]{cite}
\else
  \usepackage{cite}
\fi
\ifCLASSINFOpdf
\else
\fi
\begin{document}
%
\title{Towards Comprehensively Understanding the Run-time Security of Programmable Logic Controllers: A 3-year Empirical Study}
%
%
%
%

\author{Rongkuan Ma,
        Qiang Wei,
        Jingyi Wang,
        Shunkai Zhu,
        Shouling Ji,
        Peng Cheng,
        Yan Jia,
        Qingxian Wang
        \IEEEcompsocitemizethanks{\IEEEcompsocthanksitem Rongkuan Ma, Qiang Wei and Qingxian Wang are with the State Key Laboratory of Mathematical Engineering and Advanced Computing, Zhengzhou, Henan 450001, China. Qiang Wei is the corresponding author.
        Email: \{rongkuan307, prof\_weiqiang, wangqingxian2015\}@163.com
\IEEEcompsocthanksitem Jingyi Wang, Shunkai Zhu, Shouling Ji and Peng Cheng are with  the State Key Laboratory of
Industrial Control Technology, Zhejiang University, Hangzhou 310000, China.
 Email: \{wangjyee, shunkaiz, sji\}@zju.edu.cn, saodiseng@gmail.com
 \IEEEcompsocthanksitem Yan Jia is with the College of Cyber Science, Nankai University, Tianjin, China.  
 Email: jiay@nankai.edu.cn}%
}

\IEEEtitleabstractindextext{%
\begin{abstract}
Programmable Logic Controllers (PLCs) are the core control devices in Industrial Control Systems (ICSs), which control and monitor the underlying physical plants such as power grids.
PLCs were initially designed to work in a trusted industrial network, which however can be brittle once deployed in an Internet-facing (or penetrated) network.
Yet, there is a lack of systematic empirical analysis of the run-time security of modern real-world PLCs. 
To close this gap, we present the first large-scale measurement on 23 off-the-shelf PLCs across 13 leading vendors.
We find many common security issues and unexplored implications that should be more carefully addressed in the design and implementation. 
To sum up, the unsupervised logic applications can cause system resource/privilege abuse, which gives adversaries new means to hijack the control flow of a runtime system remotely (without exploiting memory vulnerabilities);
2) the improper access control mechanisms bring many unauthorized access implications;
3) the proprietary or semi-proprietary protocols are fragile regarding confidentiality and integrity protection of run-time data.
We empirically evaluated the corresponding attack vectors on multiple PLCs,
which demonstrates that the security implications are severe and broad. 
Our findings were reported to the related parties responsibly, and 20 bugs have been confirmed with 7 assigned CVEs.
\end{abstract}

}

\maketitle


\IEEEdisplaynontitleabstractindextext


%
\IEEEpeerreviewmaketitle

\section{Introduction}
Industrial Control Systems (ICSs), such as power systems, oil and gas transmission systems, chemical plants, and smart manufacturing, often play a vital role in critical infrastructures and people's livelihood.
Thus, attacks targeted at ICSs can potentially cause more destructive consequences \cite{langner2011stuxnet,nell2016impact}, 
such as explosions, widespread power outages, and even loss of life, than that against information technology (IT) systems.
\par
Programmable Logic Controllers (PLCs) are the core control devices in ICSs, 
which connect cyberspace and the physical world by controlling and monitoring the physical plants directly. 
The security of PLCs at run-time is of great significance in protecting an ICS from catastrophic physical damages.
For instance, in two well-known ICS-tailored malware instances \cite{langner2011stuxnet, triton2018}, 
attackers have utilized PLC vulnerabilities for damaging physical facilities.
In the Stuxnet attack, the adversary launched PLC-tailored attacks to sabotage the uranium enrichment facilities in Iran by injecting malicious logic code into a Siemens PLC \cite{langner2011stuxnet}. 
The TRITON implanted a more advanced privileged backdoor in the Triconex MP3008 controller \cite{triton2018}.
The myth of the air-gapped ICS network has been overthrown over the past decade \cite{fivemyths}, and well-skilled attackers are proven to be able to penetrate a ``closed'' network. 
Thus, it is expected that PLCs themselves should be designed to be sufficiently secure instead of solely relying on network isolation or firewalls.
\par
Prior works have disclosed different kinds of vulnerabilities that could be exploited by attackers,
such as hardcode passwords \cite{wardak2016plc}, 
firmware modification \cite{basnight2013firmware,garcia2017hey}, 
memory corruption \cite{tfuzz2011plc,muench2018you},
and flaws in non-encryption protocols \cite{beresford2011exploiting, GHALEB201862}.
These works are often inspired by conventional security implications that are investigated on a few legacy PLCs. Yet, Modern PLCs have taken considerable and progressive measures to address these issues. 
\textit{Systematic characterization and understanding of the run-time security of recent modern PLCs are thus urgently needed.}
Specifically, the implication of unsupervised logic applications at run-time has not been adequately studied before, which however could bring tremendous threats according to our study.
The access control (AC) and authentication mechanisms are also far from providing effective protection. 
Besides, the improved communication protocols still lack confidentiality and integrity protection of run-time sensitive information.
While existing works only focus on a few dedicated PLCs, 
their methodologies and insights lack sufficient generality. 
In conclusion, the community still lacks a comprehensive understanding of the run-time security of currently deployed leading PLCs.
\par
In this paper, 
we close this gap by presenting the first large-scale and systematic security measurement of recent real-world PLCs developed by popular vendors.
We focus on the run-time security of PLCs since it is the main operating state in their lifecycle with stronger protection measures in place (compared to compile-time). 
Performing such a security measurement on PLCs is extremely challenging for the following notorious reasons.
First, all popular PLCs are closed-source
and the protocols supported by them are proprietary with almost no publicly available technical details.
Furthermore, dynamic analysis methodologies are difficult to apply for PLCs due to the well-known limitations of instrumentation techniques and firmware emulators \cite{muench2018you}.
Specifically, since logic application binaries are usually in proprietary formats, existing binary analysis techniques are difficult to apply.
The diversity of the runtime systems, logic application formats, and industrial protocols, further hinders a large-scale security analysis of real-world PLCs.
\par
To address these challenges,
we present simplified yet unified methodologies that are effective to evaluate the following three classes of security implications at a PLC's run-time.
\par
\textit{a) Unsupervised logic applications.}
Logic application binaries are interpreted and executed by a PLC's runtime system.
We find that the execution of logic applications often has not been properly supervised by PLCs.
To utilize such a design flaw, we successfully craft a novel attack called One-time Cost Attack, which can even hijack the execution flow of the runtime system by crafting a malicious logic application. This attack exploits the remotely programmable attribute of PLCs and does not need complex memory vulnerability exploitation techniques.
\par
\textit{b) Coarse-grained and vulnerable access control mechanisms.}
A PLC needs to support multiple types of users for rich-featured operations \cite{61131stand}.
However, we find that PLCs often over-trust workstations in practice, which results in many violations of the requirement that 
``all entities should be identified and authenticated for all access to the control system'' in the IEC 62443‑4-2 standard\cite{62443}.
In our study,
we identified many unauthorized privileged access vulnerabilities caused by coarse-grained AC mechanisms and vulnerable authentication processes.
\par
\textit{c) Insecure industrial protocols.}
Various industrial protocols are designed and implemented by PLC vendors 
complying with the IEC-61131 standard,
which inherently lacks security consideration. Some leading vendors have made efforts to enhance their proprietary protocols' security \cite{lei2017spear,biham2019rogue7}.
However, according to our study on the popular PLCs with the latest firmware, we find that they are still far from protecting sensitive information at run-time.
\par
To demonstrate the impacts of the above security implications, 
we conducted a large-scale measurement on 23 PLCs from 13 different vendors, 
7 of which are among the top 20 global automation vendors \cite{top50vendors}, and find that most of them are affected.
We further dig into the underlying causes behind these vulnerabilities, which PLC vendors and the standard-makers should carefully address to improve the secure design and implementation of future PLCs. 
In summary, we make the following contributions.
\begin{itemize}
\item We conduct a systematic analysis of the run-time security of 23 recent leading PLCs, uncover new design flaws, and propose new attack methods exploiting them.
\item We demonstrate the threats and consequences of our findings with end-to-end Proof-of-Concept (PoC) attacks against real-world PLCs, e.g., remotely launching a one-time cost attack to get the root access of a PLC, or bypassing their password authentication mechanism to manipulate a controller arbitrarily.
\item We conduct a large-scale measurement on 23 PLCs from 13 leading industrial automation vendors.
We find that 14 of them are subject to over-permissive logic application implications, which can be exploited to hijack the control flow of the runtime system remotely.
All the evaluated PLCs lack critical data protection, e.g., unauthenticated data access to 5 vendors' PLCs by bypassing the same vulnerable authentication process.
We have reported all the discovered security issues to the related parties responsibly.
Multiple zero-day vulnerabilities and security notifications have been disclosed by vendors to alarm and mitigate these risks with 7 assigned CVEs. 
\item We plan to release all the traffic data and the automatic security analysis details in this study to the community to benchmark and facilitate future studies in this area.
\end{itemize}
\section{Related Work}
\label{sec:related}
\textbf{Logic Application Modification Attacks.}
In the literature,
researchers have studied logic application modification attacks,
which can deploy PLC-tailored and physics-aware payloads to destruct a control system \cite{sun2020sok}.
Specifically, 
\cite{govil2017ladder}, \cite{timebomb2011s7} discussed how logic bombs could interrupt a control process stealthily.
\cite{spenneberg2016plc} implemented a worm using the PLC programming language to infect other PLCs from one PLC. 
\cite{mclaughlin2011dynamic}, \cite{mclaughlin2014trusted} and \cite{zhang2019towards} 
conducted a series of studies on automated malicious code generation and the safety vetting of industrial control processes.
\cite{keliris2018icsref} studied how to automatically generate malicious logic code in the compiled binaries
and \cite{kalle2019clik} further studied how to hide the malicious code in victim PLCs.
In addition, \cite{abbasi2016stealth}, \cite{garcia2017hey} studied how a rootkit works in the PLC, 
which could damage the physical plants and is hard to detect.
In contrast, our study finds the security implication of unsupervised logic application brought by PLCs' design, and we present three kinds of attack vectors to demonstrate the threats and conduct a large-scale measurement on 23 PLCs.
\par
\noindent
\textbf{PLC Vulnerabilities.}
For legacy PLCs, 
several industry reports showed a set of vulnerabilities,
including a lack of basic authentication and authorization \cite{beresford2011exploiting,milinkovic2012industrial,inherencevul},
lack of encryption in communication protocols \cite{basecamp},
firmware modification or backdoors \cite{basnight2013firmware,schuett2014evaluation,ma2019stealthy}.
Existing analysis of modern PLC vulnerabilities mainly focuses on protocols and software vulnerabilities.
For protocols, 
researchers have demonstrated flaws in S7comm \cite{beresford2011exploiting}, \cite{GHALEB201862}, \cite{wardak2016plc}, EthernetIP/CIP \cite{senthivel2017scada}, Codesys protocols \cite{inherencevul} 
and their updated versions  \cite{klick2015internet}, \cite{lei2017spear}, \cite{biham2019rogue7},
involving plaintext protocols, 
predictable session ID, and unauthorized sensitive control command. 
For software vulnerability, 
\cite{tfuzz2011plc}, \cite{muench2018you}, \cite{zaddach2014avatar}, and \cite{costin2014large} 
studied fuzzing and binary analysis methodologies to identify PLC software vulnerabilities (e.g., buffer overflow).
The exposed vulnerabilities are conventional software flaws and these studies lack a systematic understanding of PLCs' features.
In sharp contrast,
we conduct a comprehensive study on the run-time security issues of PLCs by design, from the perspectives of run-time code security and data protection. 
\par
\noindent
\textbf{Empirical Study on Other Systems Bugs.}
Many works have been conducted to study bugs in different types of systems. 
Quarta et al. \cite{quarta2017experimental} analyzed unique bugs in the robotics domain from a systems security standpoint.
Fernandes et al. \cite{fernandes2016security} and Zhou et al. \cite{zhou2019discovering} studied security hazards on emerging smart home platforms regarding the interactions between IoT devices, mobile Apps, and backend clouds.
Zhao et al. \cite{zhao2020large} performed a ten-month-long empirical study on the vulnerability of 1 362 906 IoT devices to reveal the primary security threats that IoT devices face and the defenses that vendors and users adopt.
Ning et al. \cite{ning2019understanding} exposed a new attacking surface that universally exists in ARM-based platforms (i.e., development boards, IoT devices, cloud servers, and mobile devices).
Unlike this prior work, we perform an empirical study in a different domain, i.e., the run-time security of PLCs (the core devices in ICSs).
\section{Background}
\label{sec:bcd}
\subsection{PLC Ecosystem}
\begin{figure}[htbp]
  \graphicspath{{./}}
  \centerline{\includegraphics[width=0.42\paperwidth]{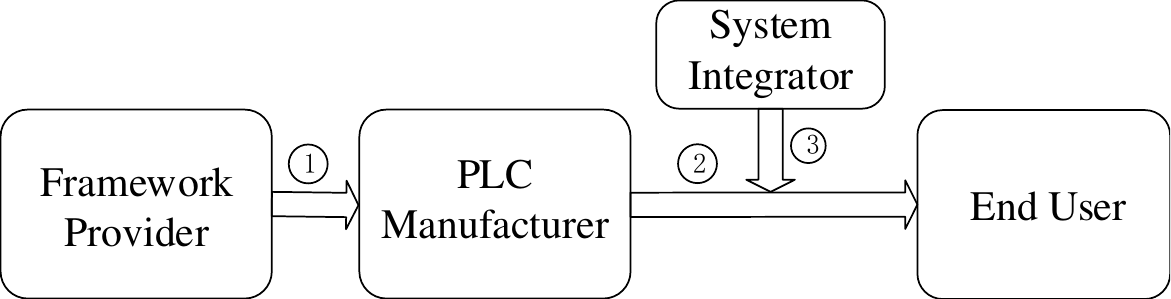}}
  \caption{The PLC Ecosystem.}
   \label{eco}
\end{figure}
Fig.~\ref{eco} shows the typical PLC ecosystem. Framework providers provide basic programming frameworks (e.g., 3S Smart Software Codesys \cite{codesys}) for PLC Manufacturers.
A programming framework includes the frontend (i.e., programming software) and backend (i.e., runtime system).
Part of PLC manufacturers (e.g., Siemens and Rockwell) develop their proprietary programming frameworks, while most customize it based on a basic framework provided by other vendors (e.g., 3S Smart Software Codesys).
PLC manufacturers produce PLC devices for end-users or system integrators who provide the end-users the industrial automation solutions and develop logic applications in PLCs to build an ICS (e.g., Distributed Control Systems (DCS) \cite{icswiki} for thermal power plants).
Note that the roles in the ecosystem may overlap. 
For example, Siemens and Rockwell also play the role of system integrators to provide industrial automation solutions for end-users.

\subsection{PLC Life Cycle and Operating Modes}
We divide a PLC's life cycle into \textit{compile-time} and \textit{run-time} according to our domain knowledge.
We define a PLC at compile-time before it executes the logic application following the operation manual.
At run-time, the PLC executes the application to control and monitor the physical plants.
\par 
The user can set a specific operating mode to define a PLC’s current capabilities. 
Specifically, the user can set the PLC in the Program mode when it needs to be programmed. 
Under this mode, the user can download logic applications to the PLC, and run or stop the PLC. 
In this paper, we call a PLC, which is in modes that allow developers to edit or update the logic application, operating in compile-time.
\par
At other times, the PLC operates at run-time, when it can be in various modes that define multiple-level protections.
Under a specific run-time mode, the PLC can disable some functionalities. For example, when the Siemens CPU317 PLC operates under the write protection mode, the downloading logic application functionality is disabled. 
In this paper, we focus on the security of PLCs at run-time because in the real world PLCs in the industry generally work in this mode.

\begin{figure}[htbp]
  \graphicspath{{./}}
\centerline{\includegraphics[width=0.4\paperwidth]{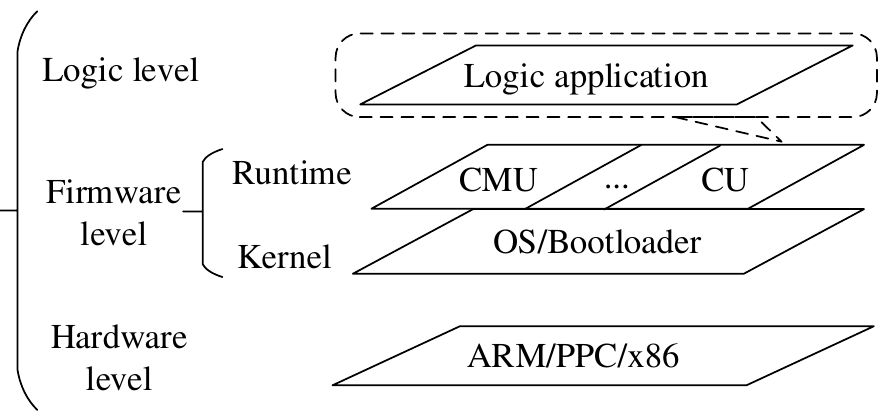}}
  \caption{Abstract hierarchical model of the PLC architecture.}
  \label{plchierarchical}
\end{figure}

\begin{figure}[htbp]
  \graphicspath{{./}}
\centerline{\includegraphics[width=0.4\paperwidth]{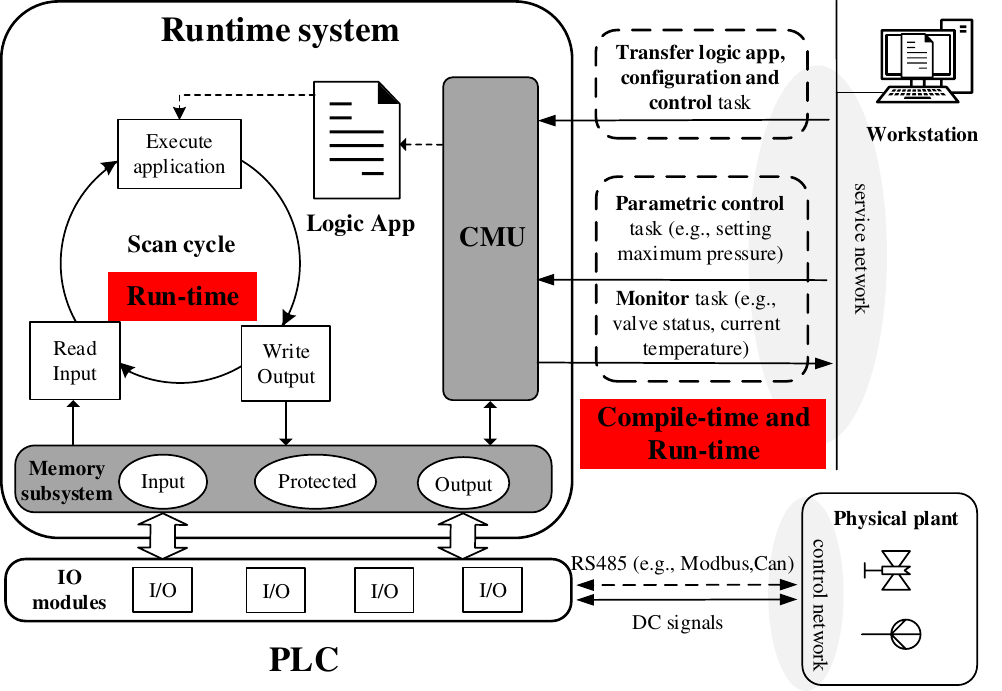}}
  \caption{The abstract functional modules of the PLC implementation.}
  \label{plcmodel}
\end{figure}

\subsection{PLC Architecture and Functional Modules}
\textbf{PLC Architecture.} As shown in Fig.~\ref{plchierarchical}, the implementation of a PLC can be generally divided into three abstract levels.
Logic applications run at the logic level and are supervised by the runtime system.
The firmware acts as a kind of operating system (OS) that may be general-purpose OS-based, embedded OS-based, or bare-metal. 
It is responsible for the interchange of the values to and from the PLC’s hardware IO modules that directly interact with the physical system. It supports communication functionality and executing and supervising logic applications through a customized runtime system.
An embedded board with a specific Instruction Set Architecture (ISA) (e.g., ARM, PPC, MIPS) can be chosen at the hardware level.  
\par
PLCs are the core devices in ICSs to control physical plants and interact with remote users using industrial protocols.
As shown in Fig.~\ref{plcmodel}, when a PLC operates at compile-time, the users can program the PLC remotely and transfer the logic applications through its communication services.
At run-time, the logic applications process the readings from and forward the outputs to the physical appliances (e.g., sensors and actuators) using I/O modules.
The input sensor readings are converted into digital values and stored in system memory. 
Meanwhile, the PLC periodically sends sensor readings to connected workstations.
Note that the communication unit (CMU) can be integrated into one firmware with other units or implemented separately in different chips.
Either way, the data flow in Fig.~\ref{plcmodel} is consistent.
\par
\textbf{Logic Application.} 
Receiving and executing external logic applications are the featured functionalities of PLCs.
At compile-time, logic applications could be downloaded to or uploaded from the PLC by the workstation.
Once the logic application is applied, the real-world ICS would operate for several years. 
The logic application does not need to update frequently.
\par
At run-time, as shown in Fig.~\ref{plcmodel}, the logic application processes input values from the input memory image and store output values in the output memory image. 
This process is known as a \textit{scan cycle}.
The execution of logic applications is supervised by the runtime system.
For example, the runtime system limits the scan cycle within a specified time.
\par
Logic applications are written in specific programming languages complying with the IEC-61131-3 standard,
such as Ladder Logic, Structured Text Language, and Function Block Diagrams.
The source code of a logic application is edited and compiled on the development software by operational technology (OT) engineers. 
The binaries of a logic application are usually in a customized and undocumented format (different across different programming software tools). The interpretation and execution of logic application binaries are also closed-source and diverse for different runtime systems. 
\par
\begin{table}[]
  \caption{The proprietary or semi-proprietary protocols and code formats of PLCs} 
  \center
  \scalebox{0.9}{
  \begin{tabular}{cccc}
  \hline
  \textbf{Vendor} & \textbf{PLC Model} & \textbf{Proprietary Protocol} & \textbf{Code Format} \\ \hline
  \multirow{3}{*}{Siemens} & CPU317 &S7comm  & MC7 \\
   & CPU1217c &S7comm-plus P3  & N/A \\
   & CPU1511-1 & S7comm-plus P3 & N/A \\
  \multirow{2}{*}{Rockwell} & ControlLogix 5561 & PCCC & N/A \\
   & Micrologix 1100 & PCCC-plus &N/A  \\
  \multirow{5}{*}{\begin{tabular}[c]{@{}c@{}}Schneider \\ Electric\end{tabular}} & M218 &M218-Codesys v3  & ARM \\
   & M241 & M241-Codesys v3 & ARM \\
   & M258 & M258-Codesys v3 & ARM \\
   & M340 &M340-Modbus  & ARM \\
   & M580 &M580-Modbus  & ARM \\
  GE & RX3i &GE-SRTP  & N/A \\
  WAGO & PFC200 &WAGO-Codesys v2  & ARM \\
  \multirow{2}{*}{Nandaauto} & NA300 & Self-owned & ARM \\
   & NA400 & Self-owned &x86  \\
  \multirow{3}{*}{Hollysys} & LK207 &Codesys v2  & ARM \\
   & LK210 &Codesys v2  & ARMBE \\
   & FM802 &Codesys v2  &x86  \\
  Triconex & MP3008 &TRISTATION  & PPC \\
  ABB & PM573 &ABB-Codesys v2  & PPC \\
  Mitsubishi & R08CPU &MelSoft  & N/A \\
  Omron & CS1H/CS1W &FINS & N/A \\
  Haiwell & T16S0P & Self-owned & N/A \\
  SIFANG & CSC850 & ProConOS & PPC \\ \hline
  \end{tabular}
  }
  \label{basicinfo}
\end{table}

\textbf{Runtime System.} 
The runtime system plays a crucial role in the PLC's firmware. 
It is responsible for 1) executing logic applications periodically by its computing unit (CU)(see Fig.~\ref{plcmodel}); 2) providing communication services for remote users and devices by its CMU (see Fig.~\ref{plcmodel});
Specifically, the runtime system also needs to satisfy the special functionality requirements of a PLC. 
For instance, it should allow a user to remotely modify the variable value in PLCs' output image through the communication services, which could immediately change the state of a valve. 
3) last but not least, the runtime system should ensure the security and safety of the execution of logic applications since it is in practice the supervisor of logic applications.
\par
The implementations of runtime systems from different vendors are diverse and mostly closed-source.
For instance, the CU in S7-300 PLC could execute the MC7 code, a specific bytecode format designed by Siemens,
while the CU in WAGO PLC could execute the native machine code compiled by its proprietary integrated development environment.
We show the logic application code formats of typical PLCs in Tab.~\ref{basicinfo}. 
\par
\textbf{Industrial Protocols.}
PLCs could support multiple industrial network protocols through the CMU based on an Ethernet network or serial buses (legacy PLCs in particular).
Our study focuses on Ethernet-based industrial protocols.
As shown in Tab.~\ref{basicinfo}, each PLC has its proprietary protocol (e.g., S7comm) or semi-proprietary protocol (e.g., M218-Codesys V3). These protocols support several mission-critical PLC functionalities, as shown in Fig.~\ref{plcmodel}, including transferring logic applications and configurations (between the PLC and workstations), modifying the variable values in PLC memory by sending parametric control commands, and remotely diagnosing the PLC status.
\par
Many vendors implement their security-enhanced proprietary or semi-proprietary protocols because standard protocols like Modbus and common industrial protocol (CIP) cannot meet their requirements in practice.
For example,
encryption and authentication are not considered in the standard Modbus and CIP protocols by design.
Schneider and Rockwell have implemented relatively secure protocols based on them. As shown in Tab.~\ref{basicinfo}, the M340-Modbus protocol is a secure protocol based on the Modbus, and the PCCC protocol is a secure protocol based on the CIP protocol.
These proprietary and semi-proprietary protocols usually support more privileged functionalities than standard protocols, but their security has not been analyzed sufficiently.
Note that PLCs also support general-purpose protocols in IT practices (e.g., HTTP, FTP, Telnet), which are however not enabled by default due to the resource constraint of embedded controllers. 

\section{Overview of PLC Security Analysis}

\begin{table*}[htbp]
  \centering
  \caption{Overview of PLC security problems}\label{tab1} 
  \scalebox{1.0}{ 
    \begin{tabular}{|l|l|l|}
        \hline
        \multicolumn{1}{|c|}{\textbf{\begin{tabular}[c]{@{}c@{}}Security \\Implications\end{tabular}}}&\multicolumn{1}{c|}{\textbf{Problems}}&\multicolumn{1}{c|}{\textbf{Description of the Corresponding Attacks}} \\
        \hline
\multirow{3}*{\begin{tabular}[c]{@{}c@{}}Unsupervised \\Logic App\end{tabular}}&Over-permissive instructions & Hijacking the runtime’s control flow \\
        \cline{2-3}
        &Unsupervised illegal code &Crashing the running firmware \\
        \cline{2-3}
        &System resources abuse &Dead-loop logic code attack \\
        \hline
\multirow{2}*{\begin{tabular}[c]{@{}c@{}}Risky AC\end{tabular}}&Coarse-grained AC Mechanism&Rogue workstation attack against unauthorized access \\
        \cline{2-3}
        &Vulnerable authentication process&Authentication bypass for privileged manipulations \\
        \hline
\multirow{3}*{\begin{tabular}[c]{@{}c@{}}Insecure Comm.\end{tabular}}&Weak encryption&Sniffing attack \\
        \cline{2-3}
        &\multirow{2}*{Lack of sensitive data integrity protection} & Spoofing workstations attack\\
        \cline{3-3}
        & &False Data Injection (FDI) attack\\
        \hline
    \end{tabular}}
    \label{overview}
\end{table*}

\begin{table*}[htbp]
  \caption{The statistical summary of the attacks affecting the evaluated PLCs}\label{tab2} 
  \center
  \sbox\tempbox{%
  \scalebox{1.0}{
  \begin{tabular}{cccc|cccc|ccc}
  \hline
  \multirow{3}{*}{\begin{tabular}[c]{@{}c@{}}Attack\\ Vector\end{tabular}} & \multicolumn{3}{c|}{Unsupervision} & \multicolumn{4}{c|}{ Unauthorized Privileged Access}      & \multicolumn{3}{c}{Insecure Communication }   \\ 
  \cline{2-11} 
  &\begin{tabular}[c]{@{}c@{}}Over-permissive\\ Instructions\end{tabular} & \begin{tabular}[c]{@{}c@{}}Illegal\\ Code\end{tabular} & \begin{tabular}[c]{@{}c@{}}Resource \\Abuse  \end{tabular} & \begin{tabular}[c]{@{}c@{}}Upload\\ App\end{tabular} & \begin{tabular}[c]{@{}c@{}}Read/Write\\ Vars\end{tabular} & \begin{tabular}[c]{@{}c@{}}Run/Stop\end{tabular} & \begin{tabular}[c]{@{}c@{}}Download\\ App\end{tabular}  & Sniff &Spoof& FDI \\ \cline{1-1}
  \hline
  \hline
 \begin{tabular}[c]{c}Affected\\ PLCs\end{tabular}  &3 & 14/23 & 6/23  & (8+8)/19  & (7+8)/19 & (6+8)/19 & (3+7)/19 & 21/21  & 19/21  & 19/21 \\ 
  \hline
  \end{tabular} }
  }
  \label{summary}
  \setlength\templen{\wd\tempbox}
  \usebox{\tempbox}\\[3pt]
  \parbox{\the\templen}{\small 
  The ``Upload App" represents uploading logic applications from PLCs;
  the "Read/Write" Vars represents reading/writing variables from/to PLCs;
  the "Run/Stop" represents run/stop the PLC;
  the "Download App" represents downloading logic applications to the PLC;
  the detailed description of the four attack vectors is demonstrated in Tab.~\ref{attvec}.
  The "Sniff" represent the sniffing attack; the "Spoof" represents the spoofing attack; the FDI represents the false data injection attack; the detailed description of the three attack vectors is demonstrated in Sec.~\ref{sec:com}.
 }
\end{table*}

\textbf{Threat Model and Scope.} We assume that attackers could access the network ports of the proprietary or semi-proprietary protocols of the PLCs at run-time. 
They \textit{do not} need to compromise a workstation as Stuxnet did \cite{langner2011stuxnet} or use a JTAG tool to access the PLC physically as previous work did \cite{basnight2013firmware,garcia2017hey}.
This is a reasonable assumption in practice.
As shown by the Shodan search engine \cite{shodan}, there are lots of Internet-facing PLCs, e.g., over 60000 exposed Siemens and Rockwell PLCs.
In addition, the exposed ICS attack incidents \cite{langner2011stuxnet,triton2018} have proved that attackers can penetrate a closed ICS network, and compromise PLCs directly.
\par
Our study focuses on discovering and exploiting the design flaws in the security of PLCs at run-time.
Thus, we do not consider those attacks which attempt to exploit a software vulnerability (e.g., buffer overflow and XSS vulnerabilities) in the PLC firmware. 
Those bugs can be patched. 
In contrast, attacks against design flaws have a more far-reaching impact since they are difficult to fix without a redesign.
\par
\textbf{Summary of Analysis Results.} 
We chose our benchmarks based on the following considerations: 1) they include 9 global vendors that are all ranking among the top 30 global automation vendors; 2) for some vendors (e.g., Siemens, Rockwell, Schneider), our benchmarks cover above 70\% of their PLC serials. Thus, we think they are representative. 
\par
From the perspective of code execution and data protection at PLC's run-time, we classify the found design flaws into three high-level categories, as shown in Tab.~\ref{overview}.
We demonstrate the detailed attack vectors and the statistical results of the affected PLCs in Tab.~\ref{summary}\footnote{Hereafter, a/b means a out of b, where a and b are two numbers.}.
In Sec.~\ref{sec:exe},~\ref{sec:acc}, and~\ref{sec:com}, 
we explore these security implications in detail respectively.
\par
\textbf{1) Unsupervised Logic Applications.} 
According to the specification of PLCs' featured capabilities, they need to receive and execute external code (i.e., logic applications). 
The logic applications should run in a supervised environment to ensure security and safety.
We experimentally analyze and test 23 PLCs from leading vendors.
As shown in Tab.~\ref{summary},
about 6/23 PLCs suffer from resource abuse against dead-loop logic code, and 14/23 PLCs would crash against illegal compiled code.
\par
\textbf{2) Risky Access Control Mechanism.} 
We analyze 19 PLCs of our benchmark for their AC mechanisms.
The evaluation of Schneider M218\&M241\&M258 PLCs is excluded due to their unpopular AC mechanism (detailed explanation in Sec. V).
As shown in Tab.~\ref{summary},
above half of PLCs suffer from the corresponding unauthorized privileged access attack vectors.
Specifically, four types of unauthorized privileged access can impact 8/19, 7/19, 6/19, and 3/19 PLCs operating under run-time modes respectively.
After exploiting the password authentication of the vulnerable PLCs, the number of impacted PLCs is 16/19, 15/19, 14/19, and 10/19 respectively (shown in Tab.~\ref{summary}).
\par
\textbf{3) Insecure Data Transmission.} 
We evaluated 21 PLCs against three man-in-the-middle (MITM) attacks because two PLCs were unfortunately damaged when testing the attacks in Sec.4.
PLCs still lack secure communication protocols that can protect the integrity and confidentiality of the transmitted data. 
As shown in Tab.~\ref{summary}, 
19/21 of the evaluated PLCs lack effective protection against spoofing and FDI attacks, and all PLCs lack protection against the sniffing attack.


\section{Unsupervised Logic Applications}
\label{sec:exe}

As aforementioned, in the PLC's ecosystem and life cycle, an adversary has multiple chances to deploy customized logic applications in a PLC over the Internet.
Both at compile-time and run-time, by exploiting an authentication bypass vulnerability or launching a MITM attack, an adversary can inject malicious code into a PLC, as we will detail in Sec.~\ref{sec:acc} and Sec.~\ref{sec:com}.
Thus, the PLC's runtime system, as the supervisor of logic applications, should ensure that the capabilities of the logic applications are strictly supervised (in terms of system calls, illegal code, and resource usage) to prevent bad consequences.
However, according to our dynamic analysis of runtime systems, logic applications often have over-permissive system privileges during execution, which could bring tremendous threats.
The problem has been largely overlooked in the literature so far. 
In the following, we evaluate the security risks at the execution time of logic applications by presenting three kinds of attack vectors on PLCs: 
\textit{\textbf{1) One-time Cost attack to exploit over-permissive instructions execution to hijack the control flow of the runtime system.}}
\textit{\textbf{2) Unsupervised illegal code execution to crash the running firmware.}} 
\textit{\textbf{3) Dead-loop code injection to abuse system resources.}}
\begin{figure}[htbp]
  \graphicspath{{./}}
  \centerline{\includegraphics[width=0.40\paperwidth]{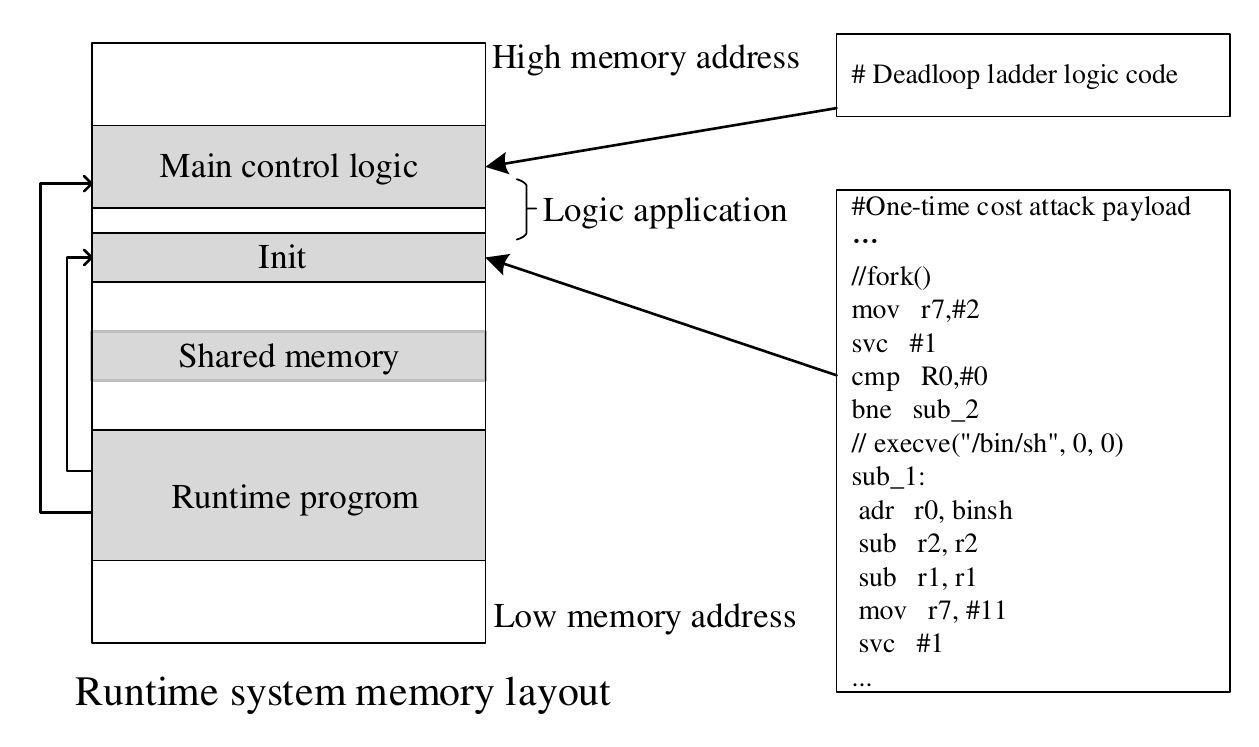}}
  \caption{Malicious code in the memory of a runtime system.}
  \label{layout}
\end{figure}

\subsection{Attack Principle}
The following demonstrates how we uncover and exploit the above PLC design flaw in practice. 
In the IEC-61131 standard, there is no sufficient specification on the implementation of how a runtime system interprets, executes, and supervises logic applications.
Thus, we have to manually reverse-engineer several PLCs' runtime systems to understand the execution of logic applications better. 
Specifically, we find that the binary code of a logic application is loaded in an executable memory space of the runtime system process, and more importantly, executed without effective supervision.
As a consequence, an adversary can customize a logic application (as shown in Fig.~\ref{layout}) to execute over-permissive instructions and unsupervised illegal code, or abuse system resources at run-time.
\par
\textbf{One-time Cost Attack.} 
As mentioned in Sec.~\ref{sec:bcd}, logic applications are written in specific programming languages by programming toolchains that are implemented by complying with the IEC-61131-3 standard, and running on the top-level layer. Therefore, to be secure, only predefined language instructions and library functions can be legitimately used in logic applications. If the instructions are not properly limited by the supervisor (the runtime system), a malicious logic application can execute over-permissive instructions and even hijack the supervisor’s control flow to execute privileged code.
\par
We propose a novel \textit{one-time cost attack} to exploit the over-permissive instructions execution implication, which inserts a shell backdoor to the logic application as shown in Fig.~\ref{layout}.
Conducting such an attack, however, is highly non-trivial for closed-source platforms. 
First, we need to identify the logic applications from 
the packaged project file,
where logic applications are packaged with other resource files by the programming software.
Second, we need to parse the format of logic applications which is usually not supported by existing tools.
To overcome the challenges, we manually reverse-engineer the compiling process of the programming software to identify the logic applications. 
Note that this is a laborious but one-time effort.
This study focuses on logic applications with readable code instructions (e.g., disassemble instructions).
We defer how to identify proprietary immediate representation instructions of PLCs as future work.
After the identification, we customize a similar methodology to ICSREF~\cite{keliris2018icsref} to reverse-engineer the format of the logic application.
Note that to deal with logic applications of CodeSys-based PLCs, we improve the ICSREF on finding static libraries and their offsets in a more generic way.
Our changes have been submitted and accepted on the open-source \texttt{icsref} GitHub repository\footnote{https://github.com/momalab/ICSREF}.
Finally, we need to reverse-engineer the runtime system to understand how it interprets and executes the logic application in the memory space (as shown on the left of Fig.~\ref{layout}) to construct the attack payload properly. 
\par
By reverse-engineering analysis (using the existing tools IDA Pro \cite{idapro} and gdbserver \cite{gdbserver}), we find several challenges to successfully inserting a Remote Access Trojan (RAT) into a PLC. 
First, we should avoid executing the payload periodically since we need and only need to execute the shellcode once. 
Moreover, we should evade the detection of the runtime system, watchdogs, or other unknown monitoring systems. 
For example, the runtime system supervises the execution of the logic application to guarantee that it runs within a specified time. 
If it exceeds the time limit, the runtime system or watchdog will perform certain special operations to stop it. 
Thus, we take two steps to address the challenges using several features of the logic application:
1) Deploy the attack payload in the initial subroutine of the logic application because it is only executed once before the first execution cycle; 
2) Invoke the fork system call to create a new process and jump to the normal logic application function later, which could avoid aborting the normal execution of the logic application and thus can evade detections of watchdogs and timing monitors.
\par
\textbf{Unsupervised illegal code.} Similarly, the runtime system should also supervise the execution of the illegal code to ensure the safety of the running firmware because the illegal code can interrupt 
the execution flow of the running firmware.
We take two steps to test whether the runtime system can gracefully handle the inserted illegal code in a logic application.
First, we determine whether their low-level representation code format is native machine code. 
We utilize an open-source tool (i.e., \texttt{binwalk} \cite{binwalk}) to identify it. 
Then, to evaluate whether a PLC is affected by the illegal code, we construct some simplified illegal machine codes embedded in the logic application to download to a PLC. 
For example, we replace the legal instruction 7C 08 02 A6 (mflr r0) with illegal machine code FF FF FF FF (illegal instruction) to test the ABB PM573 PLC. 
In this way, a PLC executing illegal code without proper sanitization measures would crash.
\par
\textbf{Dead-loop Logic Code.} As mentioned, to meet the real-time requirements of the ICS environment, the thread of logic applications in a PLC usually has the highest priority. 
Consequently, if there is no proper resource limitation, maliciously crafted logic applications could occupy too many resources of the PLC to trigger a DoS attack. 
Fig~\ref{layout} shows an example where we insert dead-loop ladder logic code (e.g., infinite loop and recursive call) into the logic application.
If not properly scheduled by the runtime system, the dead-loop logic code can abuse the system resource. As a result, the system will fall into an infinite loop and then reject to respond to other requests.
\subsection{Evaluation}
We conduct a large-scale study on 23 PLCs from leading vendors to confirm whether the PLCs suffer from the unsupervised logic application security implication. 
Respectively, we conduct three kinds of attacks: one-time cost attack against three PLCs, illegal code execution against 15 PLCs, and dead-loop logic injection against 23 PLCs.
\begin{table*}[ht]
  \hspace{0.5cm}
  \caption{The evaluation of unsupervised logic application implication in the PLCs}
  \begin{center}
  \sbox\tempbox{%
  \scalebox{0.9}{
  \begin{tabular}{|p{2.5cm}<{\centering}|p{3cm}<{\centering}|p{2cm}<{\centering}|p{2.5cm}<{\centering}|p{3cm}<{\centering}|c|p{2.5cm}<{\centering}|}
  \hline
   \textbf{Vendor}  &\textbf{Model }&\textbf{Check}&\textbf{Dead-loop }&\textbf{Illegal Code }&\textbf{\begin{tabular}[c]{@{}c@{}}Vendor \\ Response\end{tabular}}&\textbf{\begin{tabular}[c]{@{}c@{}}Over-permissive \\ Instruction\end{tabular}}\\
  \hline
  \multirow{3}*{ Siemens}& CPU317 &monitoring &halt &--& &\\
  \cline{2-7}
    & CPU1217c &monitoring &halt &--& &\\
  \cline{2-7}
    & CPU1511-1 &compile-time&halt &--& &\\
  \hline
  \multirow{2}*{Rockwell}& ControlLogix 5561 &monitoring &halt &--& &\\
  \cline{2-7}
    & Micrologix 1100 &monitoring &halt &--& &\\
  \hline
  \multirow{5}*{Schneider}& M218 &monitoring &halt &crash\&no recovery& confirmed&\\
  \cline{2-7}
    & M241 &monitoring &halt &crash& confirmed&\\
  \cline{2-7}
    & M258 &monitoring &halt &crash& confirmed&\\
  \cline{2-7}
    & M340 &monitoring &halt &crash& confirmed&\\
  \cline{2-7}
    & M580 &monitoring &halt &crash& confirmed&\\
  \hline
  GE& RX3i &\xmark &DoS &crash&  &\\
  \hline
  WAGO&PFC200&monitoring &halt&crash& confirmed &One-time Cost\\
  \hline
  \multirow{2}*{Nandaauto}& NA300 &monitoring &DoS (High CPU) &crash& confirmed&One-time Cost\\
  \cline{2-7}
    & NA400 &monitoring &halt&crash& confirmed&\\
  \hline
  \multirow{3}*{ Hollysys}& LK207 &monitoring &reboot &crash &&\\
  \cline{2-7}
    & LK210 &monitoring &reboot &crash &&One-time Cost\\
  \cline{2-7}
    & FM802 &monitoring &reboot &crash &&\\
  \hline
  Triconex& MP3008 & compile-time&N/A&N/A &&\\
  \hline
  ABB& PM573 &monitoring &halt &crash &confirmed&\\
  \hline
  Mitsubishi& R08CPU &\xmark &halt &-- &&\\
  \hline
  Omron& CS1H/CS1W &monitoring &halt &-- &&\\
  \hline
  Haiwell&T16S0P&compile-time &N/A &-- &&\\
  \hline
  SIFANG&CSC850 &\xmark &DoS &crash\&no recovery &confirmed&\\
  \hline
  \end{tabular}}
  }
  \setlength\templen{\wd\tempbox}
  \usebox{\tempbox}\\[3pt]
  \parbox{\the\templen}{\small 
  * -- means that we do not identify what it is.
  The ``Check'' column means how a Dead-loop logic is handled. 
  ``monitoring'' means the workstation could monitor abnormal behaviors. \xmark means the PLC does not handle the abnormal behavior.
  }
  \label{isolationres}
  \end{center}
\end{table*}
\par
\textbf{1) Over-permissive Instructions.} 
It is laborious to deploy the \textit{one-time cost attack} against various PLCs case by case.
We conduct three case studies of the one-time cost attack on WAGO PFC200 PLC, Hollysys LK210 PLC, and Nandaauto NA300 PLC. 
The three PLCs adopt different runtime systems and hardware. 
\par
\textbf{Case Study.} 
In the case study, we demonstrate the details of the attack PoC on the WAGO PFC200 PLC.
The WAGO PFC200 PLC adopts the ARMv7 processor hardware platform, the Linux-based operating system, and the Codesys \cite{codesys} based runtime system \textit{plclinux\_rt}.
\par
Based on the previous work ICSREF, 
we identify the init subroutine (MEMORY\_INIT) in the compiled logic application.
To conduct our attack, we need to craft the native machine code to trigger several system calls. 
To avoid interrupting the attack by the supervisor (the runtime system), we should avoid using unsupported instructions and syscalls, and triggering unknown exceptions.
\par
We find that the subroutines in the logic application are loaded into a section of memory image with the \textit{RWX} property.
The MEMORY\_INIT is only executed once before the first \textit{scan cycle}. 
Therefore, we replace a section of less critical code in the MEMORY\_INIT function with our shellcode.
This could ensure our shellcode is executed only once when the logic application is loaded and executed.
After the crafted logic application is downloaded in the PLC, 
a tcp\_reverse command shell connects to the attacker's host in our experiment, which inherits the Linux root privilege of the runtime system.
We also provide a video to show this attack process on YouTube\footnote{https://youtu.be/sfFaw\_WfxRM}.
\par
\textbf{2) Unsupervised illegal code execution.} 
In our experiment, 
\textit{14/23 PLCs are confirmed supporting native machine code logic applications (see Tab.~\ref{basicinfo}) and could be crashed by our simplified illegal codes (see Tab.~\ref{isolationres}).}
To ensure that the logic applications do not disappear after power off, PLCs need to store them in non-volatile random-access memory, such as flash memory.
In general, we can use the capabilities of the PLC's proprietary protocol to download the illegal codes to the volatile random-access memory or write them in flash memory.
Therefore, if the application with illegal code is stored in flash memory, the PLC would enter a DoS state with no recovery. 
The reason is that the illegal code could not be cleared after rebooting and would be loaded and executed once the PLC powered on.
An adversary could utilize this feature to cause more harmful effects on PLCs. 
During our evaluation, the M218 PLC is damaged by this issue and has to be returned to the manufacturer for repair, which could bring unacceptable loss in a real industrial environment.
We have reported our findings to related vendors as potential remote code execution vulnerabilities.
Most of them have confirmed the threats (see Tab. \ref{isolationres}).
\par 
\textbf{Discussion.} 
The PLC should refuse to accept and execute the over-permissive instructions and illegal codes because the executed over-permissive instructions can hijack the execution flow of the runtime system, and the illegal codes can interrupt the execution flow of the running firmware. To fix this problem, the PLC can verify whether the instructions in a logic application are legal and safe by using promising methods such as formal methods.
\par
Note that the other PLCs that support vague logic application instructions are not ensured immune to the unsupervised illegal code execution risk because of our conservative and lower-bound analysis.
We leave the analysis to accurately recognize the instructions and formats of logic applications in future work. 
Similarly, since the PLCs do not strictly supervise the execution of illegal codes (shown in Tab.~\ref{isolationres}), they have potential over-permissive instruction implications. They are potentially vulnerable to the one-time cost attack.
\par 
To launch the one-time cost attack, attackers need to know about the format of the PLC logic application and its execution mechanism, which is usually not well publicly documented. It took us about 36 hours to acquire this knowledge of WAGO PFC200 PLC and Hollysys LK210 PLC and about 50 hours to Nandaauto NA300 PLC by reverse engineering. The WAGO PFC200 PLC and Hollysys LK210 PLC adopt Codesys-based runtime system with different versions and boards with different Instructions Set Architectures. Thus, we can use ICSREF to speed up the analysis process. 
With these one-time manual efforts, the attacker could easily craft a one-time cost attack.
Because the implementation details of different PLCs vary, an attacker needs to do a similar analysis on PLCs of new types. 
Nevertheless, the manual effort could be eased with the development of automatic tools (e.g., ICSREF).
\par
\begin{figure}[t]
  \graphicspath{{./}}
  \centerline{\includegraphics[width=0.4\paperwidth]{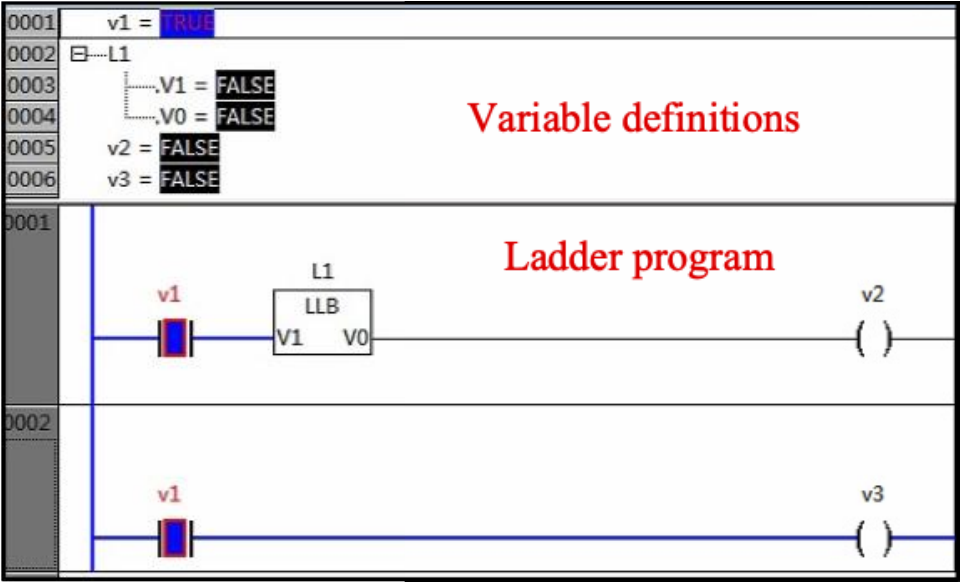}}
  \caption{Improved dead-loop test case.}
  \label{dl}
\end{figure}
\textbf{3) Dead-loop Logic Code.}
We conduct a large-scale study to evaluate the security implication of abusing system resources in PLCs.
Similar to the above evaluation, we should avoid writing the dead-loop code into the flash memory.
For example, the CSC850 PLC was damaged by the dead-loop code similarly to the M218 PLC.
In addition, we design a safer test case to perform the evaluation.
In these cases, the PLC could be recovered even though we have to write the logic application in its non-volatile memory by its proprietary protocol.
As shown in Fig. \ref{dl}, the L1 block is the dead-loop block. 
When v1 is forced as ON (0x01), the dead-loop code is triggered. 
After rebooting the controller, v1 would be initialized as OFF (0x00) (i.e., in the init process); thus the dead-loop code would not be triggered. 
\par
As shown in Tab.~\ref{isolationres}, \textit{6/23 PLCs cannot handle the dead-loop code gracefully and would behave as DoS or rebooting when executing the dead-loop code.}
In other PLCs, 
the dead-loop code could cause a ``halt'' error, 
which could be monitored by the workstation and recovered online.
We think this is still risky in practice because it could affect the Input/Output of the PLC, as demonstrated by previous works \cite{timebomb2011s7}\cite{govil2017ladder}. 
The checking measures should be taken before executing or at the compile-time to
eliminate these implications.
However, according to our study, 
only three programming software of the PLCs (Siemens CPU1511, Triconex MP3008, and Haiwell T16S0P PLC) check the simple infinite loop at compile-time, among which only the Triconex MP3008 PLC checks the infinite recursive call.
\par
\textbf{\textit{Remarks:}} 1) The unsupervised logic application implication widely exists in popular PLCs by design, which could lead to system DoS and even arbitrary privileged code execution. 
2) In some PLCs, the resource occupied by the logic application is not scheduled properly, which could also be exploited by an adversary to cause DoS or even more severe device damage. 

\section{Risky Access Control Mechanism}
\label{sec:acc}
The IEC-62443 standard states that all entities should be identified and authenticated for all accesses \cite{62443}.
However, vendors do not sufficiently achieve this goal in practice.
According to our preliminary investigation,
about 30\% of disclosed PLC vulnerabilities (in the past 10 years) from 9 leading vendors are caused by improper AC \cite{cvedetails}. 
Despite vendors' progressive measures to fix these issues and enhance PLCs' AC security, the latest PLCs still suffer from the following two design flaws: 1) coarse-grained access control that PLCs authorize the remote users improperly at run-time; 2) vulnerable authentication process that PLCs overtrust their workstations resulting in brittle and ineffective authentication.

\subsection{Methodology}

PLCs have adopted different access constraints for different operating modes in the PLC's lifecycle, whereas a PLC in run-time modes often has more strict constraints than compile-time mode.
To achieve this, vendors may use different implementations, 
such as a hardware knob or a connection password.
In the following, 
we present how we identify the flaws in PLCs' AC mechanisms, leveraging which we uncover multiple types of novel vulnerabilities. 

\begin{figure}[t]
  \graphicspath{{./}}
\centerline{\includegraphics[width=0.4\paperwidth]{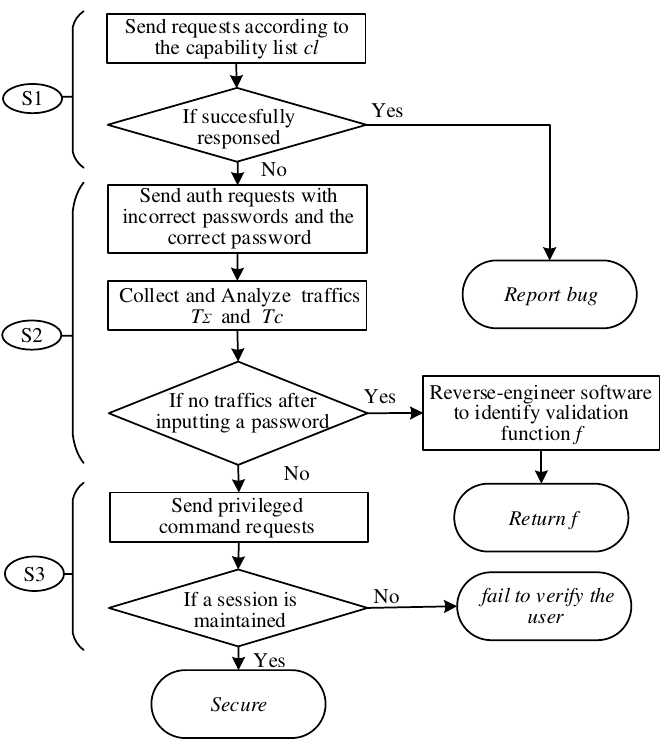}}
  \caption{Analysis of the AC Process.}
  \label{ACprocess}
\end{figure}

\begin{figure*}[htbp]
  \subfigure[Proper authentication process.]{
    \begin{minipage}{5.5 cm}
      \includegraphics[height=5cm, width=2.3 in]{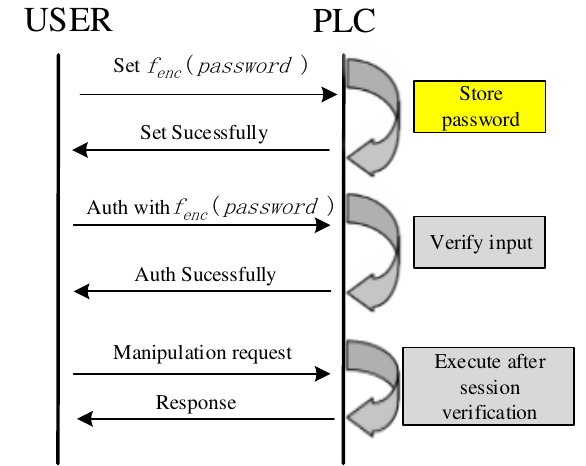}
    \end{minipage}
    \label{fig:acs}
    }
  \subfigure[Vulnerable password authentication]{
       \begin{minipage}{5.6cm}
       \includegraphics[height=5cm, width=2.3 in]{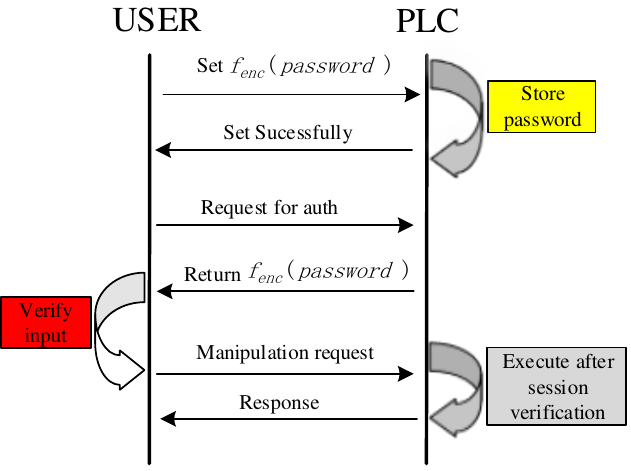}              
       \end{minipage}
       \label{fig:acv1}
       }%
  \subfigure[Failed authenticated user verification.]{
        \begin{minipage}{5.6cm}
        \includegraphics[height=5cm, width=2.5 in]{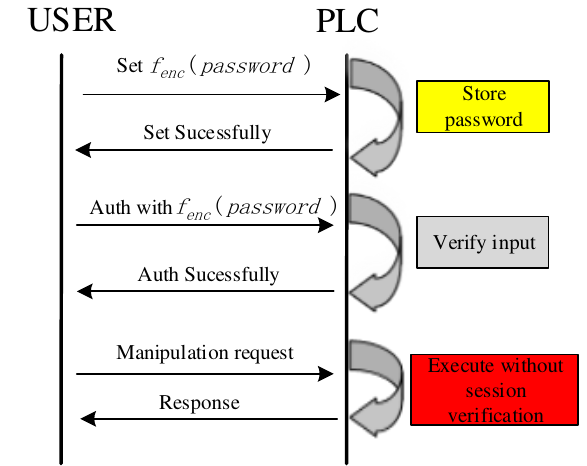}              
        \end{minipage}
        \label{fig:acv2}}%
  \caption{Authentication processes.}      
  \label{acvul}                                      
\end{figure*}
\par
\textbf{Analysis of the AC Process.} 
To start the analysis, we have to understand how an AC mechanism is deployed in a specific PLC. 
This is non-trivial due to the incomplete documentation and the closed-source platform (i.e., the programming software and protocol). 
Fig.~\ref{ACprocess} shows our analysis procedure of the PLC AC process.
In stage 1 (S1 in Fig.~\ref{ACprocess}), in the ideal case, we first need to uncover the capability list $cl$ (a list of allowed manipulations) for each mode in the PLCs. 
It is however infeasible to enumerate the whole list. 
Our remedy is to probe several fundamental but security-sensitive requests on each mode to test whether the protection is sufficient.
In stage 2, we need to uncover the authentication process of a PLC. 
Beforehand, we show a securely designed password authentication process in Fig.~\ref{fig:acs} to understand the uncovered design flaws better. 
Note that we assume a password has been successfully set in the PLC and focus on the authentication process as follows. 
The user first attempts to authenticate by sending an input password to the PLC. 
Then, \textit{a) the PLC verifies the input password and returns success if it is identical to the one set in the PLC and denies otherwise.} 
If the authentication is successful, the user can send a privileged request to the PLC. 
\textit{b) The PLC executes the request after verifying that it is from the same user}. 
The above parts are the two key rules for secure password authentication, which are often found violated in real-world implementations based on our analysis. 
\par
S2 and S3 in Fig.~\ref{ACprocess} show the analysis procedure of the authentication process.
The idea is to simulate the authentication process by inputting passwords, followed by traffic analysis and software reverse-engineering. 
We first input incorrect passwords and the correct password to generate traffics. $T_{\Sigma}$ and $T_c$ respectively denote the traffic generated from a workstation to a PLC after inputting incorrect passwords and the correct password. 
By traffic analysis, if there are no additional packets are generated for password validation. 
\textit{The implication is that the password validation is completed on the client side (programming software) instead of the PLC.} 
We then apply reverse engineering to locate the password validation function $f$ in the software by manually following the data and control flow of the corresponding subroutines. 
If traffics is generated after a password is inputted, we conjecture that the validation is finished on the PLC side. 
Then, we aim to analyze further whether the PLC verifies the user before executing a privileged request. 
We first run differential analysis on $T_c$ and $T_{\Sigma}$ and 
reverse-engineer the programming software to understand how the workstation and PLC interact in the password authentication process. 
Then, we test whether the following privileged connection session is maintained properly and return \textit{fail to verify the user} if the answer is no. 
Note that this problem can only occur if the process is implemented in a stateless protocol.
A stateless protocol means that a server does not retain the information of a session or the status of every communicating partner in multiple requests \cite{stateless}. 
Otherwise, the authentication is considered as \textit{secure} which adheres to the design in Fig.~\ref{fig:acs}.    




\textbf{Rogue Workstations.} 
We set up a rogue workstation to craft control commands and send them to the PLC. 
Since we cannot fully simulate a proprietary protocol, 
we use its programming software to assist our analysis. 
We design the following three sets of tests to evaluate the security risks in the AC mechanism which can be exploited.
\textit{\textbf{1) Capability List Probe.}} We select five mission-critical operations (which are supposed to be security-sensitive and thus need authentication), craft the commands, and send them to a PLC for every supported mode by the rogue workstation.
If a PLC does not require authentication in a specific mode, 
we consider it as a \textit{too coarse-grained protection vulnerability.} 
\textit{\textbf{2) Client-side Password Validation.}} 
If a PLC sends back the password (even if encrypted or hashed) to the client side for password validation, 
we attempt to directly modify the validation result in the binaries of the programming software. 
We use the modified software to connect and manipulate the PLC without knowing the correct password.
If successful, we consider it as an \textit{authentication bypass vulnerability}(as shown in Fig.~\ref{fig:acv1}).
\textit{\textbf{3) Authenticated User Verification.}} We simulate the protocol to send privileged control commands after a legal user's successful authentication. 
Suppose the PLC does not validate whether these commands are from the authenticated user or not and executes the privileged requests. In that case, we consider it an \textit{authentication bypass vulnerability}(as shown in Fig.~\ref{fig:acv2}).

\begin{table*}[htbp]
  \caption{The detailed description of attack vectors against PLCs}
  \label{attvec}
  \begin{center}
  \scalebox{0.95}{
  \begin{tabular}{|l|l|l|}
  \hline
  & Attack Vector & Description \\ 
  \hline
  Attack I  & Upload logic applications   & \begin{tabular}[c]{@{}l@{}}An adversary could steal intellectual property and \\   infer the structure of the physical plants by uploading logic applications from the PLC.\end{tabular}\\ 
  \hline
  Attack II & Read/Write variables        & \begin{tabular}[c]{@{}l@{}}Unauthorized read/write variables could leave an adversary to steal the run-time data of ICSs,\\   and modify critical variables maliciously in a straightforward way \\   without MITM attacks as aforementioned in Section IV.\end{tabular} \\ 
  \hline
  Attack III  & Run/Stop/Reset the PLC      & \begin{tabular}[c]{@{}l@{}}To run/stop/reset the PLC at an inappropriate time could commove the control process of ICS \\   to cause disastrous consequences.\end{tabular}   \\ 
  \hline
  Attack IV   & Download logic applications & Reprogramming the logic applications could control the whole ICS as Stuxnet did.  \\ 
  \hline
  \end{tabular}
  }
  \end{center}
\end{table*}

\begin{table*}[tbp]
  \caption{The evaluation of default design flaws of Access Control in the PLCs}
  \label{acres}
  \begin{center}
  \renewcommand\arraystretch{1.5}
  \sbox\tempbox{%
  \scalebox{0.9}{
  \begin{tabular}{|c|p{2.6cm}|p{2.1cm}|p{1.8cm}<{\centering}|p{1.8cm}<{\centering}|p{2.0cm}<{\centering}|p{1.8cm}<{\centering}|p{1.8cm}<{\centering}|}
  \hline
    \multirow{3}*{\textbf{Vendor}} &\multirow{3}*{ \textbf{Model} }&\multirow{3}*{ \textbf{\shortstack{Run-time\\ Modes}} }& \multicolumn{5}{|c|}{\textbf{Fundamental but Critical Manipulations}}\\
  \cline{4-8}
      & & &\shortstack{Read\\Id}&\shortstack{Upload\\Pro}&\shortstack{Read/Write\\Vars}&\shortstack{Run/Stop}&\shortstack{Download\\Pro}\\
  \hline
  \multirow{5}*{Siemens}&\multirow{2}*{\textcircled{1} CPU317} &W protection  &\cmark &\cmark &\cmark &\cmark & $\oslash$\\
  \cline{3-8}
    &  & R/W protection&\cmark &$\otimes$ &\cmark &\cmark &$\otimes$\\
  \cline{2-8}
  & \multirow{3}*{\shortstack{\textcircled{2} CPU1217c\\ \&\textcircled{3}  CPU1511}} & R Access &\cmark &\cmark &$\otimes$ &\cmark &$\otimes$\\
  \cline{3-8}
  &  & HMI Access&\cmark &$\otimes$ &$\otimes$ &$\otimes$ &$\otimes$\\
  \cline{3-8}
  &  & No Access&\cmark &$\otimes$ &$\otimes$ &$\otimes$ &$\otimes$\\
  \hline
  \multirow{2}*{ Rockwell}& {\textcircled{4} Micrologix1100} & RUN password &\cmark &$\oslash$ &$\oslash$ &$\oslash$ &$\otimes$\\
  \cline{2-8}
    & \textcircled{5} ControlLogix5571&  RUN &\cmark &\cmark &public tags \cmark &N/A &$\otimes$\\
  \hline
  \multirow{3}*{GE}& \multirow{3}*{\textcircled{6} RX3i} & Level Three &\cmark &\cmark &\cmark &\cmark &\cmark\\
  \cline{3-8}
  &  & Level Two &\cmark &\cmark &read only \cmark &$\otimes$ &$\otimes$\\
  \cline{3-8}
  &  & Level One &\cmark &\cmark &read only \cmark &$\otimes$ &$\otimes$\\
  \hline
  Triconex& \textcircled{7} MP3008 & RUN password &\cmark &$\oslash$ &$\oslash$ &$\oslash$ &$\otimes$\\
  \hline
  \multirow{2}*{Hollysys}&\textcircled{8} LK207\&\textcircled{9} LK210 & RUN &\cmark &\cmark &\cmark &$\otimes$ &$\otimes$\\
\cline{2-8}
  & \textcircled{10} FM802 & default &\cmark &\cmark &\cmark &\cmark &\cmark\\
\hline
  \multirow{1}*{WAGO}& \multirow{1}*{\textcircled{11} PFC200} & password &\cmark &$\otimes$ &$\otimes$ &$\otimes$ &$\otimes$\\
  \hline
  \multirow{1}*{Schneider} & \multirow{1}*{\textcircled{12}M340\&\textcircled{13}M580} & password &\cmark &$\oslash$ &$\oslash$ &$\oslash$ &$\oslash$\\
  \hline
  \multirow{2}*{  Nandaauto}& \multirow{1}*{\textcircled{14}NA300} & password &\cmark &$\oslash$ &$\oslash$ &$\oslash$ &$\oslash$\\
  \cline{2-8}
    & \multirow{1}*{\textcircled{15}NA400} & password &\cmark &$\oslash$ &$\oslash$ &$\oslash$ &$\oslash$\\
  \hline
  ABB& \textcircled{16}PM573 &  password &\cmark &$\otimes$ &$\otimes$ &$\otimes$ &$\otimes$ \\
  \hline
  \multirow{1}*{Mitsubishi}& \multirow{1}*{\textcircled{17}R08CPU} & password&\cmark &$\oslash$ &$\oslash$ &$\oslash$ &$\oslash$\\
  \hline
  \multirow{1}*{Omron}& \multirow{1}*{\shortstack{\textcircled{18}CS1H/CS1W}} &  password &\cmark  &$\otimes$ &\cmark &\cmark &\cmark\\
  \hline
  \multirow{1}*{Haiwell}&\multirow{1}*{\textcircled{19}T16S0P} & password &\cmark &$\oslash$ &$\oslash$ &$\oslash$ &$\oslash$\\
  \hline
  \end{tabular}
  }
  }
  \setlength\templen{\wd\tempbox}
  \usebox{\tempbox}\\[3pt]
  \parbox{\the\templen}{\small \cmark means that this manipulation is executed without any authentication.
  $\oslash$ means that password authentication is needed, but we bypass it to execute this manipulation.
  $\otimes$ means this manipulation request is denied, we do not bypass the authentication remotely.
  N/A means that the PLC does not support this manipulation.
  We merged the rows of the PLC models that use the same AC mechanism, such as Siemens CPU1217c PLC and Siemens CPU 1511 PLC.
  }
  \end{center}
\end{table*}
\par
\subsection{Evaluation}
First, we probe the capability list $cl$ of the tested PLCs by four classes of fundamental but security-sensitive operations (summarized in Tab. \ref{acres}).
Among them, 
Rockwell, Triconex, and Hollysys PLCs divide modes by a hardware key configured by local users.
Typically, Rockwell AB1100 PLC and Triconex PLC also provide password authentication protection in the RUN mode.
Other PLCs adopt several run-time modes (configured by remote users).
A user could change to a specific mode by password authentication.
We simulate their protocols to send the following manipulation requests to test if their protection is sufficient:
\textbf{Attack I:} Upload logic applications; \textbf{Attack II:} Read/Write variables; 
\textbf{Attack III:} Run/Stop/Reset; \textbf{Attack IV:} Download logic applications.
The detailed descriptions are shown in Tab.~\ref{attvec}.
\par
The results of the evaluated PLCs are shown in Tab. \ref{acres}. We find that: 
\textit{Against Attacks I, II, and III, 8/19, 7/19, and 6/19 PLCs respectively have no protection even under the run-time mode.}
For example, the adversary can still stop the PLCs (e.g., Siemens CPU317\&CPU1217c\&CPU1511, GE RX3i, Omron CS1H/CS1W) without any authentication.
The details are shown in Tab. \ref{acres}.
\par
We further evaluate the PLCs adopting password authentication (shown in Tab. \ref{bypassvul}). 
Among them,
7 PLCs from 5 leading vendors (e.g., Schneider M580\&M340, Rockwell Micrologix 1100, and Triconex MP3008) suffer from the same vulnerable authentication process shown in Fig.~\ref{fig:acv1}.
The detailed results are summarized in Tab.~\ref{bypassvul}. 
This shows that \textit{the vulnerability is not a coincidental programming bug but an inherent design flaw caused by the false trust assumption on the workstations.} 
Our findings have been confirmed by related parties and \textit{5 CVE/CNVD IDs have been assigned for these zero-day vulnerabilities.}
In addition,
Mitsubishi R08CPU PLC suffers from the flaw shown in Fig.~\ref{fig:acv2},
which does not verify the source of privileged requests after a successful password validation.

\textbf{\textit{Remarks:}} 1) The AC mechanism adopted by the latest PLCs still has too coarse-grained protection implications,
which allows an adversary to execute dangerous manipulations in run-time modes. 
2) Vendors have not taken a correct trust assumption on the run-time environment by design, resulting in the same vulnerable authentication process from different vendors (despite their diverse proprietary protocols).

\begin{table*}[tbp]
  \renewcommand{\arraystretch}{1.5}
  \caption{The authentication-related vulnerability of PLCs}
  \center
  \sbox\tempbox{%
  \scalebox{1.0}{
  \begin{tabular}{|c|c|c|c|c|c|}
  \hline
  \multirow{4}{*}{Vendor} & \multirow{4}{*}{Model} &  \multicolumn{4}{c|}{\begin{tabular}[c]{@{}c@{}}Authentication-related Vulnerabilities\end{tabular}} \\ \cline{3-6} 
   &  & \begin{tabular}[c]{@{}c@{}}Weak\\ password\\ Transmission\end{tabular} & CVE & \begin{tabular}[c]{@{}c@{}}Bypassed \\Authentication\end{tabular} & CVE \\ \hline
 \multirow{1}{*}{Siemens} & CPU317 & \cmark & N/A  & write protection (unpatched) & CVE-2016-9159 \\ \hline
  \multirow{1}{*}{Rockwell}  
   & Micrologix 1100 & \cmark & CVE-2020-6984 & connection password  &CVE-2020-6988  \\ \hline
  \multirow{2}{*}{Schneider} 
   & M340 & Hash & N/A& connection password &CVE-2019-6855  \\ \cline{2-6} 
   & M580 & Hash  & N/A & connection password &CVE-2019-6855 \\ \hline
  GE & RX3i & \cmark  & N/A & \xmark & -- \\ \hline
  WAGO & PFC200 & \cmark  & N/A & \xmark & -- \\ \hline
  \multirow{2}{*}{Nandaauto} & NA300  & Hash & N/A & connection password & CNVD-2019-32859 \\ \cline{2-6} 
   & NA400  & Hash & N/A & connection password & CNVD-2019-32859  \\ \hline
  \multirow{3}{*}{Hollysys} & LK207  & \cmark & N/A & \xmark &--  \\ \cline{2-6} 
   & LK210  & \cmark & N/A & \xmark  &-- \\ \cline{2-6} 
   & FM802  & \cmark & N/A & No password & N/A \\ \hline
  Triconex & MP3008  & \cmark & N/A & connection password & CVE-2020-7483  \\ \hline
  ABB & PM573 & \cmark  & N/A & \xmark & N/A \\ \hline
  Mitsubishi & R08CPU  & \cmark & N/A & remote operation password & CVE-2020-5527  \\ \hline
  Omron & CS1H/CS1W  & \cmark & N/A & \xmark & -- \\ \hline
  Haiwell & T16S0P  & Hash & N/A & connection password & CNVD-2019-32856 \\ \hline
  \end{tabular}
  }
  }
  \label{bypassvul}
  \setlength\templen{\wd\tempbox}
  \usebox{\tempbox}\\[3pt]
  \parbox{\the\templen}{\small \cmark  means that the PLC has this vulnerability and we have 
  exploit it successfully.
  \xmark means that we do not find a corresponding vulnerability.
  Hash means that the password in the traffic has been protected by some hash function.
  N/A means that we are in process of following up with the related parties to ensure they are aware of our
  findings.
  }
\end{table*}

\subsection{Case Study}
\begin{figure}[htbp]
  \graphicspath{{./}}
  \centerline{\includegraphics[width=0.4\paperwidth]{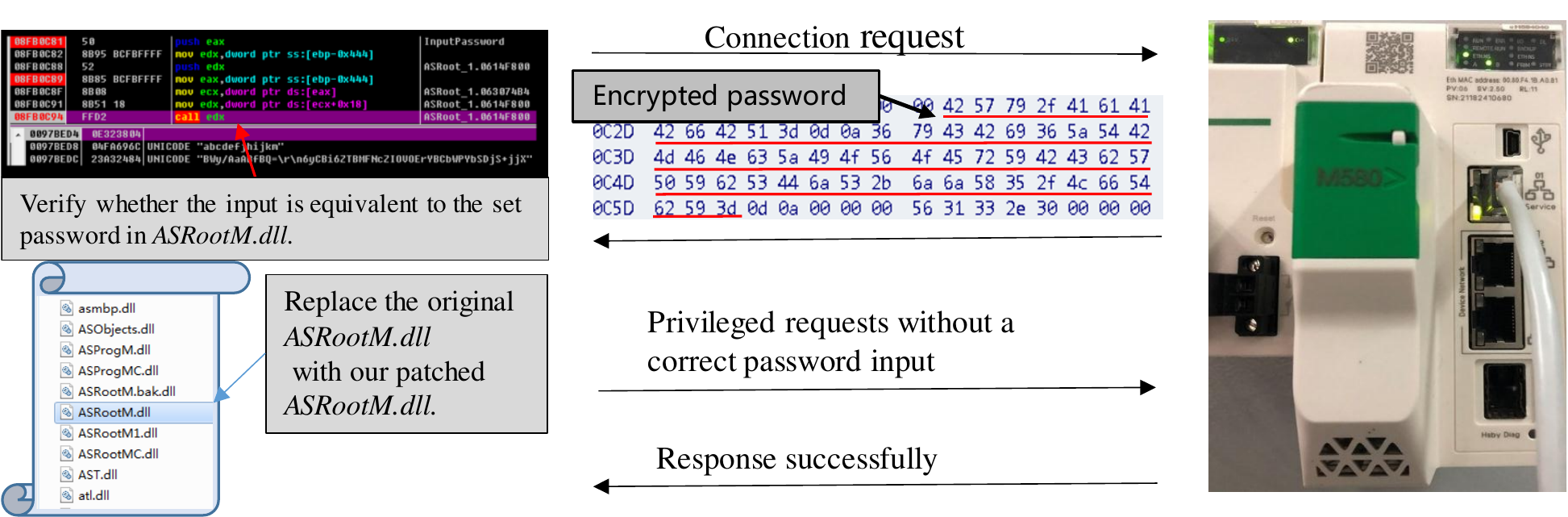}}
  \caption{Case study of M580's authentication vulnerability.}
  \label{m580}
\end{figure}
\noindent
\textbf{Coarse-grained AC Attack.} 
To demonstrate the threat of the coarse-grained AC mechanism in practice, we successfully conduct PoC attacks against PLCs from several vendors (e.g., Siemens, GE, and Schneider).
Taking the GE RX3i PLC as an example, we first set it to different protection levels.
As recommended by the official manual, the PLC runs under level four at compile-time as well as level one to three at run-time.
Then, we use a ``rogue engineering workstation'' to send the commands to read/modify an output variable value,
and find that the unauthorized workstation can read the variable value under all protection levels,
and modify the value under two run-time protection modes.
Moreover, the workstation can upload the logic program and stop the PLC, even though it is set to the highest protection level (i.e., level one).
\par
\noindent
\textbf{Authentication Bypass Exploitation.}
We conduct a case study on exploiting the authentication bypass vulnerability on the Schneider M580 PLC, Nandaauto NA300\&NA400 PLCs, Rockwell Controllogix 1100 PLC, Haiwell PLC, and Triconex MP3008 PLC.
The same vulnerable authentication process shown in Fig.~\ref{acvul}(b) can be found in these PLCs.
We demonstrate the attack against the M580 PLC as an example.
At run-time, the M580 PLC is protected by a connection password.
The remote user should authenticate for authorization before doing any privileged manipulations.
\par
We bypass the authentication vulnerability by modifying the verification result in its programming software.
In \textit{ASRootM.dll},
we modify the result of the password verification function to be constant \textit{true}, no matter what the user's input is.
As shown in Fig.~\ref{m580}, 
we replace the original \textit{ASRootM.dll} with our patched \textit{ASRootM.dll}.
Then we can download a logic application or stop the PLC without the correct password.
We also provide a video to show this attack process on YouTube\footnote{https://youtu.be/7MzA9QhzIaM}.

\section{Insecure Data Transmission}
\label{sec:com}
PLCs need to periodically send operating data to the workstation and receive control commands from the workstation at run-time.
The data leakage can help adversaries infer control models of the physical plants and then conduct concise false data injection attacks by exploiting data integrity protection flaws\cite{yang2020plc, chen2019learning, chen2020active}.
Thus, the confidentiality and integrity protection of data transmission is of vital importance. 
However, in the design of a PLC, the CMU is usually assigned with limited computing resources to meet the requirements of the logic application's execution. 
Due to the high cost of data protection algorithms (e.g., encryption and hashing), 
practical communication protocols have limited protection on the transmitted data.
This introduces security issues such as sensitive data leakage and data integrity violation. 
In this section, to evaluate such threats in real-world PLCs, 
we report a comprehensive evaluation on 18 industrial communication protocols adopted 
by 21 popular PLCs from leading vendors using three classes of MITM attacks (see Tab.~\ref{overview}).

\subsection{Methodology}
In the following, we provide the details on deploying the analysis framework regarding insecure data transmission.
This is highly non-trivial for several reasons. 
First, the communication protocols are in the binary format and closed-source. 
It is challenging to infer the format and the syntax of the protocols without human-readable information \cite{duchene2018state}.
Second, the huge traffic (hundreds of packets in a few seconds) between a PLC and a connected workstation makes manual analysis infeasible.
In addition,
the diversity of protocols (18 protocols adopted by 21 PLCs) makes it challenging to analyze with a unified methodology. 
To address these challenges, 
we propose an automated analysis framework combining reverse-engineering and differential analysis. 
Formally, we use $T$ to denote a set of packets in traffic. 
Further, let $T_S$ be the packets sent from a workstation to a PLC
and $T_R$ be the returned packets from a PLC to a workstation. 
Note that both $T_S$ and $T_R$ are the contents in the application layer 
since a protocol usually works on the application layer in the TCP/IP model. 
We first show how to automatically identify the control packet in $T_S$ 
(and respectively the returned monitoring packet in $T_R$), 
as well as the position of the critical data field in the command by reverse-engineering. 
We use $T_S$ as an illustrative example and remark that $T_R$ follows the same procedure.
\par
Our intuition is to apply reverse-engineering based on differential analysis 
to automatically identify the parametric control packet from the traffic, 
and the \textit{position} of the critical data field in it. 
We use a $\langle length, position\rangle$ pair as the result of successful identification.
The idea is that, 
we set multiple values for a certain control command 
and run each of them using the target protocol. 
From the traffic collected for each set value, 
we can filter out the parametric control packets from $T_S$ 
by locating those packets containing the set values, and 
locate the position of the critical data field by the position of the set value. 
If both the length (of the control command packets) 
and the position (of the data fields) are identical for all the set values, 
we deem a successful identification. 
Alg.~\ref{alg:gg} shows the details. 
For a target protocol $\phi$ and a specified control command $C$, 
we first generate and collect a set of packets $T_S$ with a set of constant values $X$ for the variable $v$ in $C$ (line 5 to line 8). 
Afterward, for each pair of value $x$ and its corresponding packets $T_S$, 
we filter out those packets which contain $x$ as $T'_S$ at line 11. 
Then for each packet $t$ in $T'_S$, 
we identify the length of $t$ (line 14) and the position of $x$ in $t$ (line 15). 
We add the resultant $\langle l, p\rangle$ pair into the set of candidate pairs for the current value $x$ (line 16). 
Lastly, we return the \textit{intersection} of all the possible candidate pairs for each value 
(the candidate pair that all values agree on) at line 18. 
Note that the more times we repeat setting values ($X$ with more values), 
the more certain we are on the returned result.

\begin{algorithm}[t]
  \caption{Differential Analysis}
  \label{alg:gg}
  \textbf{Input:} The target protocol $\phi$; a control command $C$; a set of constant values $X$\;
  \textbf{Output:} A set of $\langle length, position\rangle$ pairs\;
  $v\gets$ be the variable to set in $C$\;
  Let $\psi=\emptyset$ store the $\langle value,packets\rangle$ pairs\;
  \For{each value $x$ in X}{
     Set $v=x$ in $C$\;
     Run the protocol $\phi$ on $C$ and collect the packets $T_S$\;
     Add $[x,T_S]$ to $\psi$\;
   }
  Let $LP=[]$\;
   \For{each $[x, T_S]$ in $\psi$}{
     $T'_S\gets$ filter out the packets which contains value $x$ from $T_S$\;
     Let $LP_x=\emptyset$\;
     \For{each packet $t$ in $T'_S$}{
       $l\gets$ be the length of $t$\;
       $p\gets$ be the position of $x$ in $t$\;	
       Add $\langle l,p\rangle$ to $LP_x$\;
     }
     Append $LP_x$ to $LP$\;
   }
  \Return{$lp\gets$ the intersection of the elements in $LP$}\;
  \end{algorithm}

\begin{table*}[htbp]
  \renewcommand{\arraystretch}{1.1}
  \caption{The results of filtration in our traffics dataset with handcrafted traffic features}
  \center
\sbox\tempbox{%
\scalebox{1.0}{
\begin{tabular}{ccccccccc}
\hline
\multirow{3}{*}{\textbf{\begin{tabular}[c]{@{}c@{}}Traffic Dataset\\with Handcrafted Features\end{tabular}}} & \multirow{3}{*}{\textbf{Protocols}} & \multicolumn{2}{c}{\textbf{(Length, Position)}}                                    & \multicolumn{3}{c}{\textbf{Attack Vectors}}      \\  \cline{3-7} 
& & \multicolumn{1}{c|}{\multirow{2}{*}{\textbf{$T_S$}}} & \multicolumn{1}{c|}{\multirow{2}{*}{\textbf{$T_R$}}}&\multicolumn{1}{c|}{\multirow{2}{*}{\textbf{Sniffing}}} & \multicolumn{1}{c|}{\multirow{2}{*}{\textbf{Spoofing}}}&\multicolumn{1}{c}{\multirow{2}{*}{\textbf{FDI}}}  \\ 
& & \multicolumn{1}{c|}{}& \multicolumn{1}{c|}{}& \multicolumn{1}{c|}{}& \multicolumn{1}{c|}{}& \\ \hline

\multicolumn{1}{c|}{\multirow{18}{*}{\begin{tabular}[c]{@{}c@{}}$3$ times parametric control \\ commands with constant values\\ separately, continuous monitoring \\ commands with corresponding\\ responses\end{tabular}}}                       
& GE-SRTP  & (76,74) & (56,44) & \cmark & \cmark &\cmark          \\
\multicolumn{1}{c|}{}                                                                         & M241-Codesys v3    & (96,94) & (272,270) & \cmark & \cmark &\cmark         \\
\multicolumn{1}{c|}{}                                                                       
& M258-Codesys v3    & (124,82) & (176,58) & \cmark & \cmark &\cmark          \\
\multicolumn{1}{c|}{}                                                                         & M340-Modbus        & (46,37)  & (22,13)  & \cmark & \cmark &\cmark           \\
\multicolumn{1}{c|}{}                                                                         & M580-Modbus        & (46,37)  & (22,13)  & \cmark & \cmark &\cmark           \\
\multicolumn{1}{c|}{}                                                                         & MelSoft            & (89,85)  & (93,85)  & \cmark & \cmark &\cmark           \\
\multicolumn{1}{c|}{}                                                                         & FINS               & (20,18)  & (17,15)  & \cmark & \cmark &\cmark           \\
\multicolumn{1}{c|}{}                                                                         & S7comm             & (71,69)  & (55,53),(79,77)  & \cmark & \cmark &\cmark    \\
\multicolumn{1}{c|}{}                                                                         & S7comm-plus P3     & (153,124) & (225,185) & \cmark & \xmark  &\xmark         \\
\multicolumn{1}{c|}{}                                                                         & PCCC               & (71,69)   & (70,62)   & \cmark & \cmark  &\cmark           \\
\multicolumn{1}{c|}{}                                                                       
& PCCC-plus          & (99,71)  & (433,96)   & \cmark & \xmark  &\xmark          \\
\multicolumn{1}{c|}{}                                                                         & WAGO-Codesys v2    & (42,40)  & (79,73)    & \cmark & \cmark  &\cmark           \\
\multicolumn{1}{c|}{}                                                                         & ABB-Codesys v2     & (24,22)  & (19,17)    & \cmark & \cmark  &\cmark           \\
\multicolumn{1}{c|}{}                                                                         & Haiwell self-owned & (12,10)  & (12,10)    & \cmark & \cmark  &\cmark           \\
\multicolumn{1}{c|}{}                                                                         & NA300 self-owned   & (16,12)  & (16,12),(571,297) & \cmark  &\cmark &\cmark   \\
\multicolumn{1}{c|}{}                                                                         & NA400 self-owned   & (16,12)  & (16,12),(639,357) & \cmark  & \cmark &\cmark  \\
\multicolumn{1}{c|}{}                                                                         & TRISTATION         & (30,24)  & (42,24) & \cmark  & \cmark  &\cmark           \\
\multicolumn{1}{c|}{}                                                                         & Hollysys-Codesys v2 & (24,22) & (19,17) & \cmark  & \cmark  &\cmark            \\ 
\hline
\end{tabular}
}}
\label{ipr}
\setlength\templen{\wd\tempbox}
\usebox{\tempbox}\\[3pt]
\parbox{\the\templen}{\small 
  \cmark means the protocol is affected by the attack and \xmark means the protocol is resilient to the attack. Note that the S7comm-plus P3 protocol is used by the Simense CPU1217c PLC and CPU1511 PLC, and the Hollysys-Codesys v2 protocol is used by the Hollysys LK207 PLC, LK210 PLC, and FM802 PLC.
 }
\end{table*}

\textbf{Security Analysis.} Our security analysis on the protocol is based on three common kinds of MITM attacks as follows. The vulnerability of the protocols will be evidenced and evaluated by whether the attacks can succeed. For simplicity, we limit the attacks on a given control/monitoring packet. 
  \textbf{\textit{1) Sniffing attack}} succeeds if we can obtain the data value in the command from the packets.
  \textbf{\textit{2) FDI attack}} succeeds if a PLC accepts the injected fake value without integrity checking.
  \textbf{\textit{3) Spoofing attack}} succeeds if a workstation accepts the injected fake value without integrity checking.

Next, we briefly show how to deploy these attacks. 
Note that the success of Alg.~\ref{alg:gg} will immediately 
make the sniffing attack successful, and provides us with the sensitive information of the command packets (i.e., the length of the packet and the position of the data field). 
Besides, we additionally extract a \textit{signature} from the command packets ($T_S'$ in Alg.~\ref{alg:gg}) to locate the command packets in the traffic (with the same signature). 
In practice, we use a regular expression to match the signature
from the searched packets. 
The signature together with the sensitive information enables us to automatically locate the data value in the traffic and inject fake values to deploy the other two attacks. 

\subsection{Evaluation}
Following the above method, 
we evaluate the security of 18 protocols adopted by 21 PLCs from leading vendors.
We set up a testbed to collect the traffic. 
We use a CISCO switch to connect the workstation, the attacker's host, and PLCs. 15 programming software from related vendors is deployed on the workstation.
The firmware of the PLCs has been updated to the latest version.
In the reverse-engineering process, we set 3 constant values (i.e., 0x1234, 0x3456, and 0x5678) to collect the traffic between the workstation and PLCs for differential analysis. 
The collected traffics for each protocol are shown in Tab.~\ref{ipr}.
We conduct the MITM attack using open-source tools \cite{etter}. 
The detailed results are shown in Tab.~\ref{ipr} and we have the following observations.
\par
First, we can successfully reverse-engineer the sensitive information of all the evaluated protocols by locating the packets with the same $\langle length, position\rangle$ (shown in column 6 and column 7). 
We further verify the correctness of our reverse engineering by the success of the following attacks. 

\par
Second, based on the signature extracted in the filtered packets,
we are able to sniff the run-time data between a PLC and the connected workstation.
We thus conclude that \textit{all the evaluated protocols do not provide enough confidentiality on run-time data.}
Further, we evaluate the integrity protection of a target protocol against the other two kinds of attacks: \textit{a) FDI attack.}
When the workstation sends the parametric control commands to the PLC,
we replace the set value with the fake value using the tools: \texttt{ettercap} and \texttt{etterfilter} \cite{etter}.
If the PLC accepts the command with the fake data value without integrity checking,
the value will be stored in the PLC's memory as run-time data.
Using the monitoring function of the workstation,
we can verify whether the run-time data value has been modified.
If so, we confirm that the protocol is subject to the FDI attack. \textit{b) Spoofing human attack. }
We replace the returned monitoring value from the PLC to the workstation with a fake value. 
If the workstation accepts the fake value and displays it, we confirm that the protocol is subject to the spoofing human attack.
In summary, we find that \textit{16/18 protocols lack integrity checking and are subject to the above two attacks.} 
We have reported these issues to the relevant parties. 
Schneider and Mitsubishi have confirmed them as zero-day vulnerabilities: CVE-2020-7487, CVE-2020-7488, and CVE-2020-5594. 

\textbf{\textit{Remarks:}} The communication protocols adopted by latest PLCs
still have insufficient integrity checking or lack of encryption for confidentiality.
An adversary could exploit the flaws for theft, 
deception and even injecting false data to cause destructive consequences in real ICSs.


\subsection{Case Study}
\label{cs:h2}
\begin{figure}[htbp]
  \graphicspath{{./}}
  \centerline{\includegraphics[width=0.4\paperwidth]{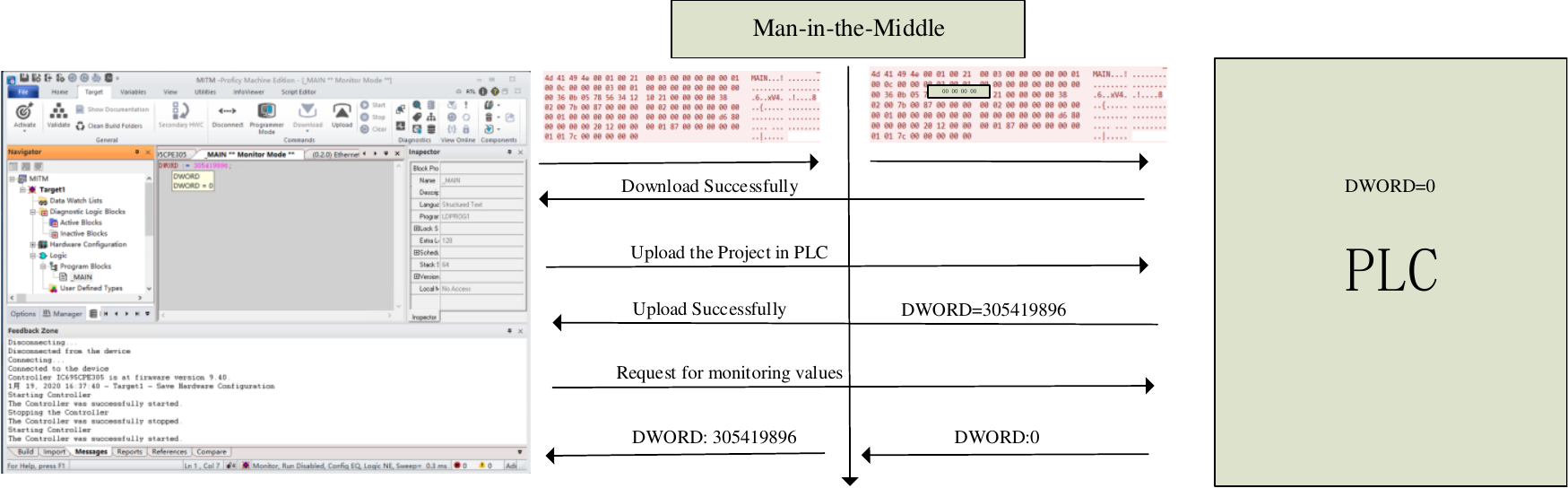}}
  \caption{An attack case study against the GE RX3i PLC.}
  \label{ge}
\end{figure}
To demonstrate the threats of insecure industrial protocols in practice, 
we conduct an attack case study against the GE (top 11 of global automation vendors) RX3i PLC
by exploiting the vulnerabilities of its proprietary protocol (i.e., GE-SRTP).
Similar attack PoCs of Schneider and Rockwell PLCs are also conducted and reported to the related vendors.
We conduct this attack as follows.
First, we simulate the normal operations on the workstation.
As shown in Fig.~\ref{ge},
we create a logic application, including an assignment statement (DWORD:=305419896),
and download the whole project to the PLC by its programming software (i.e., Proficy Machine Edition).
Then, we upload the project from PLC to simulate the process of checking whether the project has been downloaded successfully.
Finally, we use the Proficy Machine Edition to continuously monitor the run-time data in the PLC.

\par
We launch the attack in two stages as an attacker in-the-middle.
First, 
we modify the compiled code of the logic application,
replacing DWORD:=305419896 with DWORD:=0, when the project is downloaded to the PLC.
The logic application is security-sensitive and function-critical data needed to be protected strictly.
However,
the PLC accepts and executes the fake code (DWORD:=0) without any verification.
After the victim uploads the source code in the PLC and starts to monitor the run-time data,
he could see the run-time value of DWORD as 0.
The FDI attack succeeds.

In the second stage,
when the victim continuously monitors the run-time status of the PLC,
the attacker replaces the run-time data (DWORD: 0) with the crafted data (DWORD: 305419896).
Thus, the victim can see the logic code (DWORD:=305419896) with the expected run-time data (DWORD: 305419896),
while the PLC is executing the fake code (DWORD: 0).
The spoofing human attack succeeds.

In this study, 
due to the lack of proper encryption and integrity checking in the protocol,
we conduct a successful attack to 
alter the PLC's logic application by the FDI attack, meanwhile spoofing the human,
which can cause destructive effects in real ICSs.
In this process, 
a successful sniffing attack for eavesdropping the run-time data has already been leveraged.
To visualize these attacks, 
we provide a video to show the attack process on YouTube\footnote{https://www.youtube.com/watch?v=f40Jz\_\_Rwnw}.


\section{Responsible Disclosure}

We have reported all the found vulnerabilities to the related vendors.
For the vendors we cannot contact, we report their vulnerabilities to Cybersecurity and Infrastructure Security Agency (CISA) \cite{cisa} or China National Vulnerability Database (CNVD) \cite{cnvd}. 
This is a time-consuming process. We usually receive their confirmation between three months and two years since we first reported it. Until now, many reported vulnerabilities have been actively responded and addressed. We are actively following up with the remaining vendors to ensure they are aware of our findings.

\textbf{\textit{Over-permissive logic application.}} Wago and ABB have confirmed our reported vulnerabilities.
However, they think this is the normal behavior of CODESYS 2.3 runtime system, and
they cannot change the behavior of CODESYS 2.3 as end-users of this runtime solution.
ABB issued a notification to the customers on how to avoid this risk.
Schneider also has different opinions with us about this issue. 
They consider it as a protocol-related issue. 
Nevertheless, they have reported this issue in their security notification.

\textbf{\textit{Vulnerable AC.}} Schneider, Rockwell, Mitsubishi, Nandaauto and CNVD have confirmed the authentication bypass vulnerabilities we reported (some confirmed as zero-day vulnerabilities, see Tab.~\ref{bypassvul}).
Schneider, Rockwell, and Mitsubishi have published security notifications to their customers.

\textbf{\textit{Insecure data transmission.}} 
Schneider and Mitsubishi have confirmed them as zero-day vulnerabilities and fixed them in a new version. 
In particular, 
Schneider held an online meeting with us to talk about their analysis and mitigation. 
Rockwell confirmed the weak-encrypted password as a zero-day vulnerability.
However, they argue that other security issues are the inherent limitations of the CIP protocol. 
Nevertheless, they have provided a defense solution by hardening the network.
Similar to Rockwell, Siemens also claims that the S7comm is designed to work in a trusted environment.

\noindent
\textbf{Lessons.} Lessons are learned both from our findings and interactions with the vendors. 
1) The discovered design flaws in PLCs' runtime should be paid more attention and properly addressed. 
For instance, the privilege of logic applications at the execution time should be carefully limited, 
i.e., meeting the least-privilege principle to the best. 
In addition, more reliable industrial protocols should be adopted. 
This includes better-encrypted communication protocols and more strict AC mechanisms.
According to the responses we received from the leading vendors, 
\textit{vendors prefer to provide defense recommendations on the ICS network environments (e.g., firewall, IP-filtering, and sub-netting measures)
rather than improving the design of the PLC.}
These are easy yet ineffective approaches to defend against the enormous threats in the fast-growing industrial Internet-facing network.
With the advent of Industry 4.0, more and more ICSs have to open to the outside networks, thus it is more challenging to improve the security of ICSs due to PLCs' design flaws in the long term.
2) Standard-makers should enhance the security attribute of PLCs by design. 
The trust dependencies of each component and the potential security implications should be pointed out more explicitly. 
Also, more detailed specifications should be provided for ease of implementation and formal verification.   
3) Vendors should adopt a more effective vulnerability management and response system. 
Currently, 
vendors often take a long time to respond and patch for the confirmed vulnerabilities. 
For instance, 
it usually takes more than six months from our reported vulnerabilities to the completion of the repair.
Users are also not notified promptly on the software and firmware updates. 
Worse yet, some vendors charge for such updates. This leaves attackers sufficient time to exploit a discovered vulnerability.

\section{Discussion}
We did not study how PLC security would affect the end-to-end security of Supervisory Control and Data Acquisition (SCADA) systems.
Attacks against PLCs can bring devastating, infecting, and stealthy effects on the whole SCADA system. Specifically, the attacker can change the running status of physical devices (e.g., valves, pumps) by compromising the PLC. Meanwhile, the attacker can deceive the upper-layer monitoring systems in a SCADA system because they usually use the PLCs to acquire the information data of physical plants. In addition, a compromised PLC can also be a zombie computer that would be used to infect other PLCs and upper-layer systems. 
More studies should be carried out in the future to defend against ICS-tailored attacks.
\par
Our work also has limitations which could be further investigated in future work. 
1) Our study does not further inspect whether existing protection (e.g., partial encryption or hashing) adopted by some of these protocols is effective in preventing other attacks (e.g., replay attack~\cite{lei2017spear}).
We will leave these to our future research.
2) Our analysis of AC security excludes other mechanisms like the role-based AC mechanism (adopted by Schneider M218/M241/M258). 
3) Some of our analysis relies on manual or semi-automatic reverse-engineering, 
e.g., to understand the authentication process between the PLC and the workstation, 
and to identify the logic applications from the resource files. 
In the future, it is interested to develop more systematic and automatic methods for such analysis.
We are also interested in exploring how to mitigate the discovered vulnerabilities in practice. 
For instance, how to design lightweight industrial protocols balancing efficiency and security and adapting to the environment with different trustworthiness. 
Besides, we are considering more systematic methods like model checking to identify the vulnerabilities in the AC mechanisms. 
Lastly, it is interesting to evaluate further the effectiveness of different isolation methods (e.g., Sandbox, virtual machine) in practice.    



\section*{Conclusion}
We conducted the first comprehensive empirical study on the run-time security implications of PLCs regarding over-permissive logic code execution, risky AC mechanisms, and insecure data transmission respectively. 
In particular, 
we examined and measured the security of the latest PLC designs and implementations from leading vendors through a set of end-to-end attacks, 
and discovered a number of vulnerabilities.
Our research reveals the noticeable gap between existing PLCs and securely-designed ones ready for
the complicated, 
adversarial, and Internet-facing environment. 
From our findings, 
we hope that the ICS security community and vendors can take more principled approaches to improve the security in future PLC design and implementation.

\bibliographystyle{IEEEtran}
\bibliography{IEEEabrv,IEEEtrans.bib}







%



\begin{IEEEbiography}
[{\includegraphics[width=1in,height=1.25in,clip,keepaspectratio]{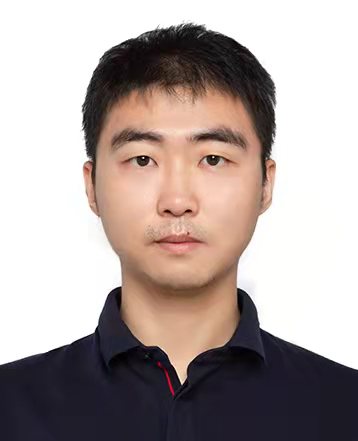}}]{Rongkuan Ma}
received his Ph.D. degree in the State Key Laboratory of Mathematical Engineering and Advanced Computing, Zhengzhou, China, in 2021. He is currently a system security researcher. His research interests include program analysis, embedded system security, iCPS security, and Web security.
\end{IEEEbiography}


\begin{IEEEbiography}[{\includegraphics[width=1in,height=1.25in,clip,keepaspectratio]{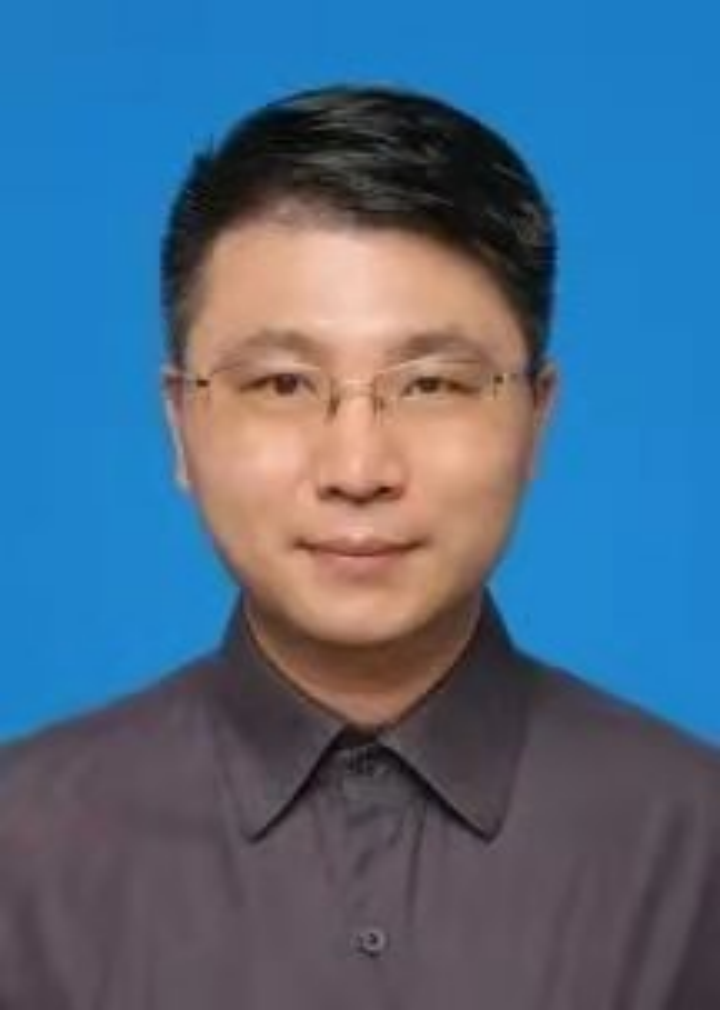}}]{Qiang Wei} received a Ph.D. degree
in Computer Science and Technology
from China National Digital Switching
System Engineering and Technological
Research Center, Zhengzhou, China. He
is currently a professor with the State Key
Laboratory of Mathematical Engineering
and Advanced Computing. His research
interests include network security, industrial
Internet security and vulnerability
discovery.
\end{IEEEbiography}

\begin{IEEEbiography}[{\includegraphics[width=1.1in,height=2.1in, clip, keepaspectratio]{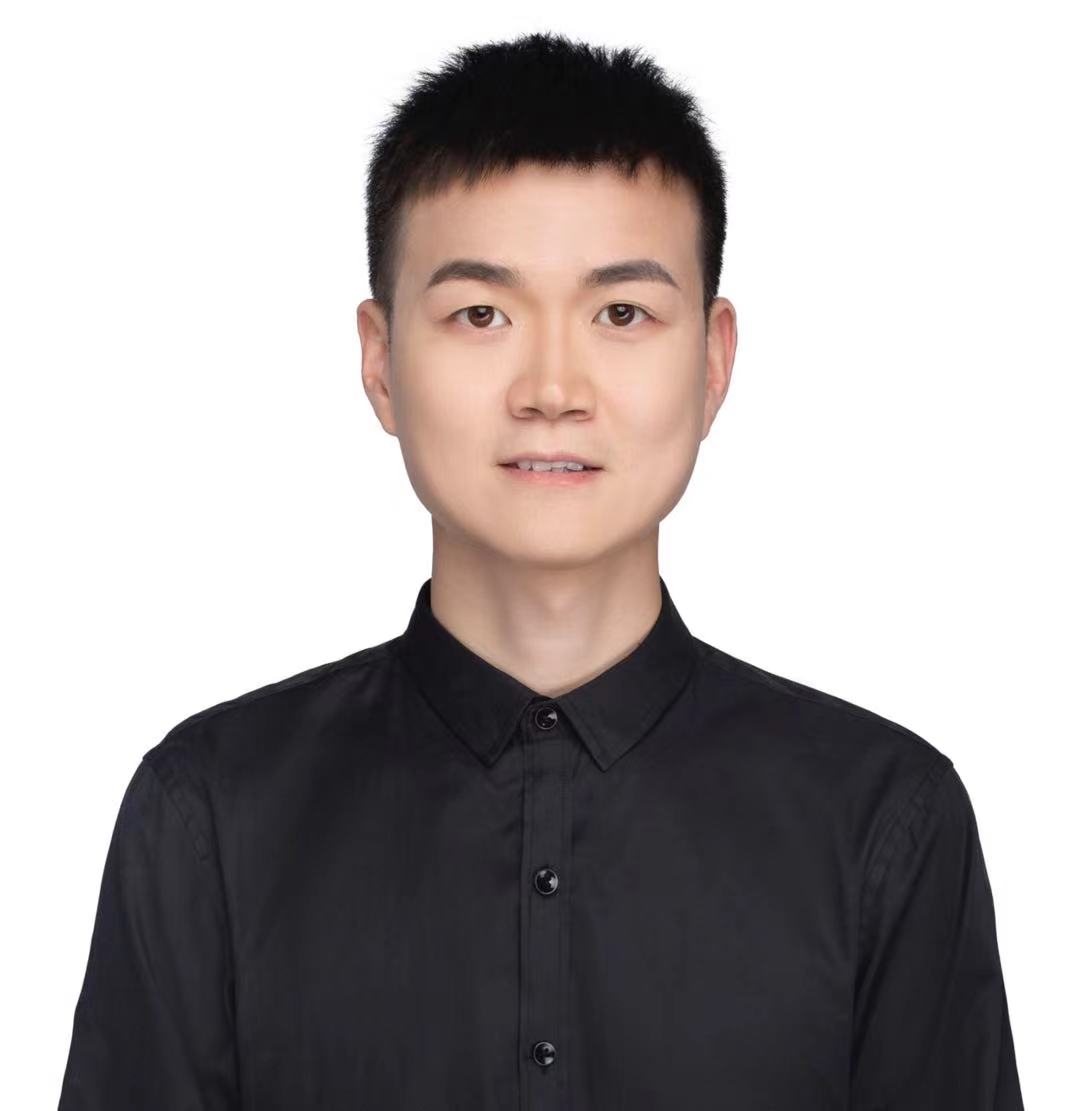}}]{Jingyi Wang} is currently a tenure-track assistant professor at the College of Control Science and Engineering, Zhejiang University, China. He received his Ph.D. from Singapore University of Technology and Design in 2018, and his bachelor’s degree in Information Engineering from Xi’an Jiaotong University in 2013. He was a research fellow at the School of Computing, National University of Singapore during 2019-2020 and at Information Systems Technology and Design Pillar, Singapore University of Technology and Design during 2018-2019. His research interests include formal methods, software engineering, cyber-security and machine learning.
\end{IEEEbiography}

\begin{IEEEbiography}[{\includegraphics[width=1.1in,height=2.1in, clip, keepaspectratio]{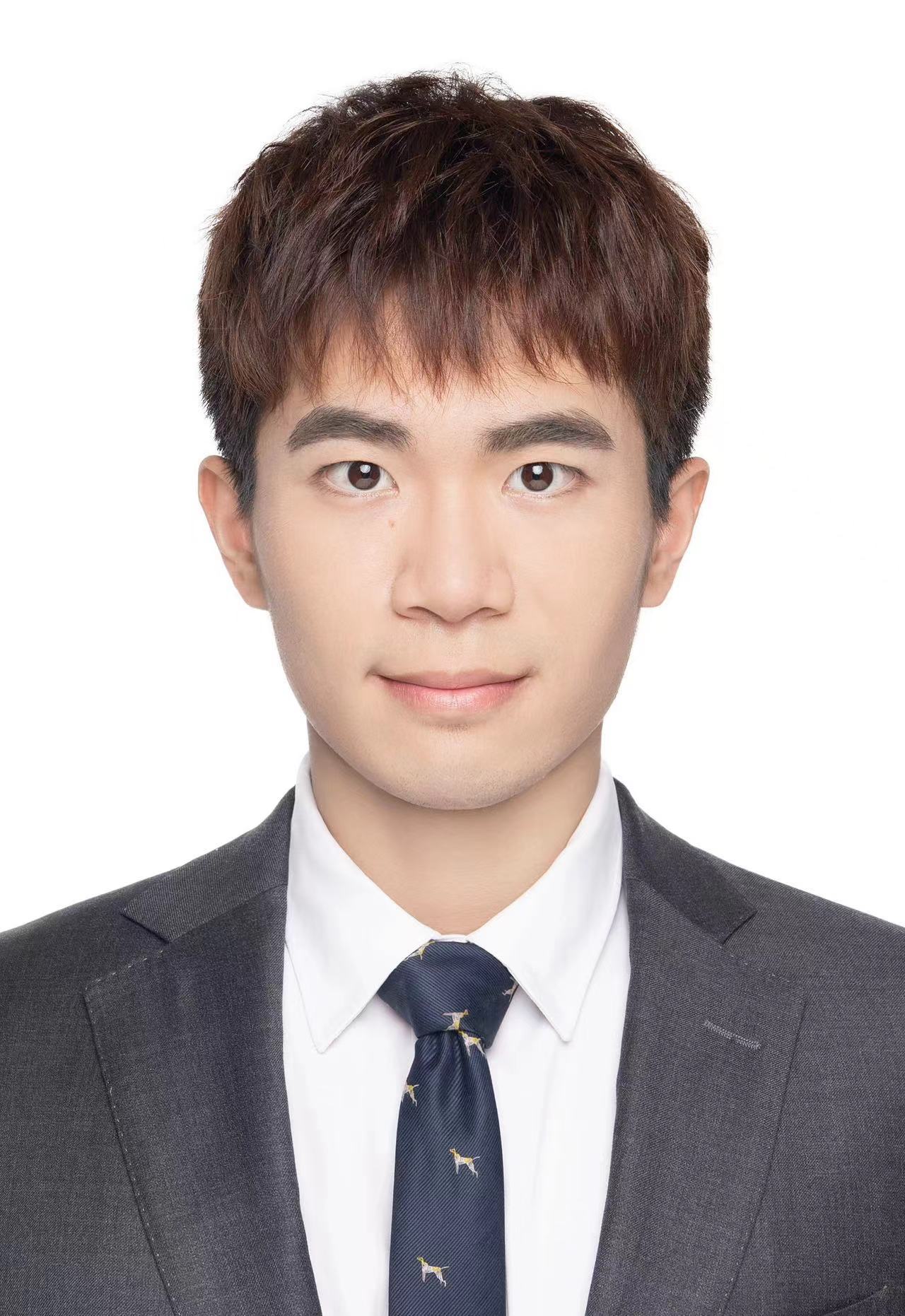}}]{Shunkai Zhu} received the B.Eng. degree in cyber security from Xidian University, in 2019. He is currently pursuing his Ph.D degree with State Key Laboratory of Industrial Control Technology, Group of Networked Sensing and Control, Zhejiang University. His research interests include software testing and control system security.
\end{IEEEbiography}

\begin{IEEEbiography}[{\includegraphics[width=1.0in,height=1.25in, clip, keepaspectratio]{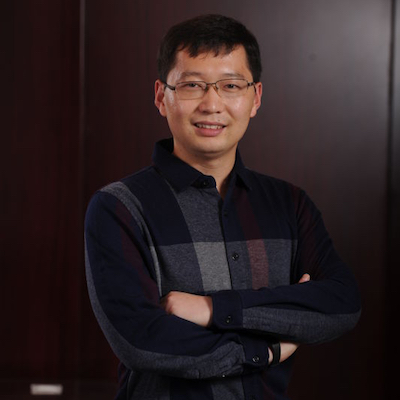}}]{Shouling Ji} is a ZJU 100-Young Professor in the College of Computer Science and Technology at Zhejiang University and a Research Faculty in the School of Electrical and Computer Engineering at Georgia Institute of Technology. He received a Ph.D. in Electrical and Computer Engineering from Georgia Institute of Technology and a Ph.D. in Computer Science from Georgia State University. His current research interests include AI Security, Data-driven Security and Data Analytics. He is a member of IEEE and ACM and was the Membership Chair of the IEEE student Branch at Georgia State (2012-2013).
\end{IEEEbiography}

\begin{IEEEbiography}[{\includegraphics[width=1in,height=1.25in,clip,keepaspectratio]{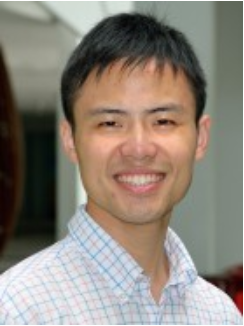}}]{Peng Cheng} (M’10) received the B.Sc. and Ph.D. degrees in control science and engineering from Zhejiang University, Hang Zhou, China, in 2004 and 2009, respectively. From 2012 to 2013, he was a Research Fellow with the Information System Technology and Design Pillar, Singapore University of Technology and Design, Singapore. He is currently a Professor with the College of Control Science and Engineering, Zhejiang University. He served as the TPC Co-Chair for IEEE IOV 2016, the Local Arrangement Co-Chair for ACM MobiHoc 2015, and the Publicity Co-Chair for IEEE MASS 2013. He serves as Associate Editors for the IEEE Transactions on Control of Network Systems, Wireless Networks, and International Journal of Communication Systems. He also serves/served as Guest Editors for the IEEE Transactions on Automatic Control and the IEEE Transactions on Signal and Information Processing over Networks. His research interests include networked sensing and control, cyber-physical systems, and control system security.
\end{IEEEbiography}

\begin{IEEEbiography}[{\includegraphics[width=1in,height=1.25in,clip,keepaspectratio]{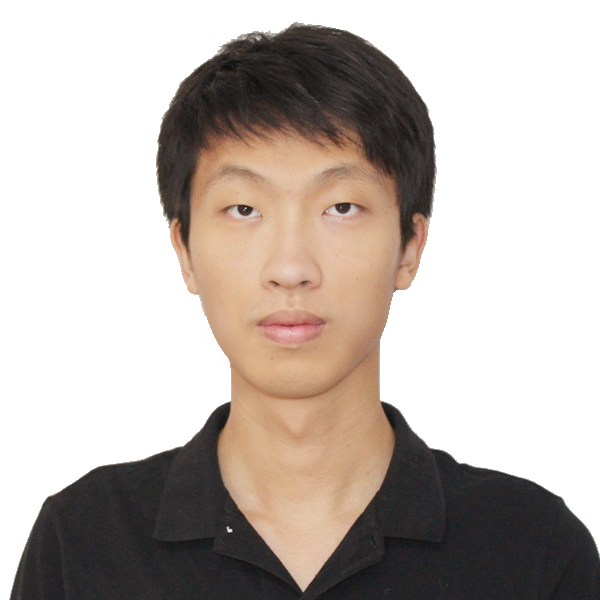}}]{Yan Jia} received the Ph.D. degree from the School of Cyber Engineering, Xidian University, Xi’an, China, in December 2020.
He is a Research Associate at the College of Cyber Science, Nankai University, Tianjin, China. His interests include discovering and understanding new design or logic security vulnerabilities in real-world systems, especially IoT systems. His work helped many high-profile vendors improve their products’ security, including Amazon, Microsoft, Apple, and Google.
\end{IEEEbiography}

\begin{IEEEbiography}[{\includegraphics[width=1in,height=1.25in,clip,keepaspectratio]{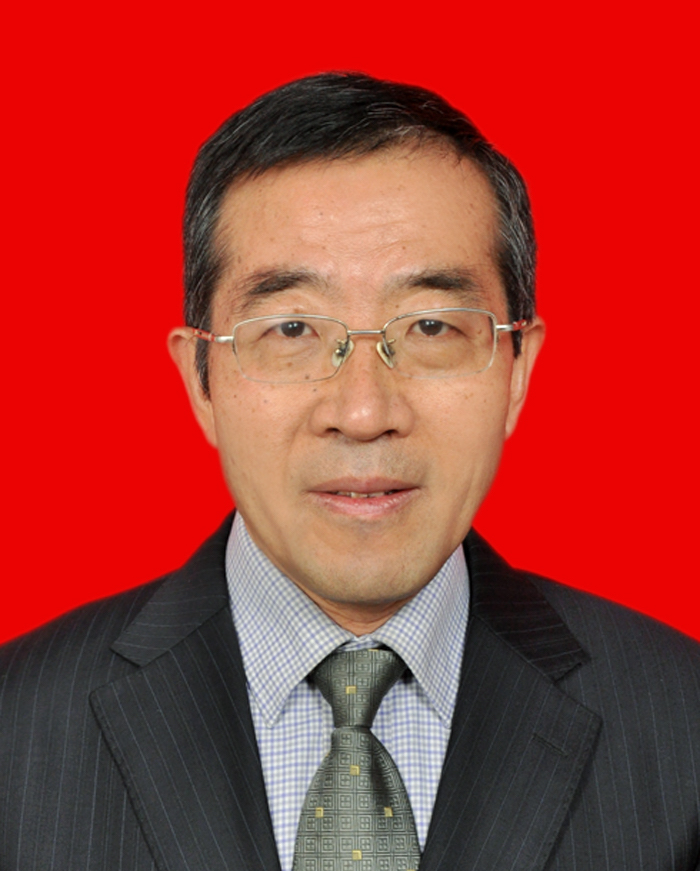}}]{Qingxian Wang} is currently a professor with the State Key Laboratory of Mathematical Engineering and Advanced Computing. His research interests include network security, industrial internet-of-things security, and vulnerability discovery.
\end{IEEEbiography}

%








\end{document}